\documentclass[preprint, aps,prl,footinbib]{revtex4-2}
\usepackage{graphicx}%
\usepackage{hyperref}
\usepackage{amsmath}
\usepackage{blindtext}
\usepackage{float}

\usepackage{xcolor}
\usepackage{amssymb}

\begin{document}

\title{Biological rhythms generated by a single activator-repressor loop with heterogeneity and diffusion.}

\author{Pablo Rojas$^1$, Oreste Piro$^{2,3}$ and Martin E. Garcia$^1$}

\affiliation{$^1$Theoretical Physics and Center for Interdisciplinary Nanostructure Science and Technology (CINSaT), University of Kassel, Kassel, Germany}
 
\affiliation{$^2$Department of Ecology and Marine Resources, Institut Mediterrani d’Estudis Avançats, IMEDEA (CSIC–UIB), Balearic Islands, Spain}
\affiliation{$^3$Departament de F{\'i}sica, Universitat de les Illes Balears, Ctra. de Valldemossa, km 7.5, Palma de Mallorca E-07122, Spain}

\begin{abstract}
Common models of circadian rhythms are constructed as compartmental reactions of well mixed biochemicals involving a negative-feedback loop containing several intermediate reaction steps in order to enable oscillations. Spatial transport of reactants is mimicked as an extra compartmental reaction step. In this letter, we show that a single activation-repression biochemical reaction pair is enough to produce sustained oscillations, if the sites of both reactions are spatially separated and molecular transport is mediated by diffusion. Our proposed scenario is the simplest possible one in terms of the participating chemical reactions and provides a conceptual basis for understanding biological oscillations and triggering  \textit{in-vitro} assays aimed at constructing minimal clocks.
\end{abstract}

\maketitle

Biological rhythms such as circadian, infradian and ultradian ones often originate in oscillations at a cellular level induced by complex mechanisms of gene auto-regulation composed by a number of coupled molecular reactions \cite{forger2017biological,gonze2002,novak2008,bharath2014}. 
An activation-repression process alone leads to a single-mode dynamical system with negative feedback unable by itself to support oscillatory behavior. 
Therefore, cellular clocks must resort to some kind of delay mechanism that effectively complements the negative feedback naturally implicit in the repression component to achieve limit-cycle dynamics. 
Most common models of such cellular rhythms are based on compartmental descriptions (individual substances reacting in a homogeneous well-mixed batch) of the biochemistry, including several intermediate molecular reactions: transcription, translation, phosphorylation, degradation, etc. 
Although the spatial inhomogeneities inside of the cell are recognized to play a role and the transport of molecules between different regions of the cell should be taken into account, this is usually done by mimicking such transport as an extra compartmental reaction \cite{goldbeter1995model,forger2017biological}. 
For example, mRNA migration from the nucleus is represented as "nuclear mRNA" reacting to produce "cytoplasmic mRNA" in the same hypothetical well mixed compartment.  In order to oscillate, these type of models need to include at least three reactions steps in the loop, but often a much larger number is required to avoid unrealistic parametrization of the reaction constants\cite{griffith68}. The purpose of this letter is to show that a much simpler reaction scheme is possible if the inhomogeneities of the reactants distributions and the transport are properly assessed. 
In particular, we will show that only one step expression-repression negative-feedback reaction is enough to produce sustained circadian oscillations, provided that the locations of transcription and translation are spatially separated and molecular transport is mediated by intra-cellular diffusion. 

In a well mixed reaction model, a single expression-repression reaction step is commonly represented in an abstract manner by the following couple of ordinary differential equations\cite{goodwin65,griffith68}
\begin{eqnarray}
\label{one-step}
\dot{M}(t)   & =& \omega_M \frac{1}{1+P(t)^h} -\gamma_M \, M(t)  \\
\dot{P}(t)  & =& \omega_P M(t) - \gamma_P \, P(t),   \nonumber
\end{eqnarray}
where $M(t)$ and $P(t)$ represent respectively the time evolution of the concentrations of the mRNA produced by a given gene and that of a protein --synthesized by the same mRNA-- that inhibits the production of further mRNA. 
The inhibitory ability of $P$ is governed by the Hill type function in the first term in the right hand side of the first equation while the second term represents the spontaneous degradation process of mRNA molecules at a rate $\alpha_M$. 
In turn, the rate of production of $P$, is proportional to the concentration $M$ and decays with a coefficient $\alpha_P$. 
The so called \textit{cooperativity} $h$ in the Hill function is a parameter that  accounts for the number of molecules necessary to inhibit the expression of the corresponding gene. 
Notice that the divergence of the vector field defining this dynamics, is constant and negative for any values of the pair $(M,P)$. 
This  means that the dynamical system contracts areas of any element of the space state and therefore it cannot support limit cycles or self-sustained oscillations \cite{griffith68}. 
Instead, it can be proved that Eqs.(\ref{one-step}) only have stable fixed points as stationary solutions. 

In the realm of compartmental models, hence, in order to produce a cellular rhythm with a biochemical circuit of expression-repression feedback such a circuit must include at least an extra intermediate reaction to introduce a delay capable of destabilizing one of these resting states. 
One such example introduced by Goodwin (\cite{goodwin65} reads as:
\begin{eqnarray}
\label{goodwin_three_steps}
\dot{M} &=&\omega_M \frac{1}{1+R^{h}}-\gamma_M M   \nonumber\\ 
\dot{P} &=&\omega_P M-\gamma_P P \label{int_step}\\ 
\dot{R} &=&\omega_R P-\gamma_R R    \nonumber
\end{eqnarray}  
where $P$ is an extra metabolite that mediates the production of the blocking protein $R$. 
This model does produce oscillations provided that $h>8$, a value quite far from a realistic interpretation so that more sophisticated models resort to include a few extra steps of phosphorylation and de-phosphorylation before the actual repression takes place\cite{gonze2002}. An extension of Eqs. (2) to $n$ reaction steps is straightforward (See Supplemental Material). 
In Fig.\ref{figure_intro} a scheme for this extension is shown, along with a plot of the minimal $h$ that produce oscillations in a compartmental model of $n$ components (values that satisfy the secant condition, first derived in \cite{tyson1978dynamics}). 
\begin{figure}[b] 
    \includegraphics[width=3.3in]{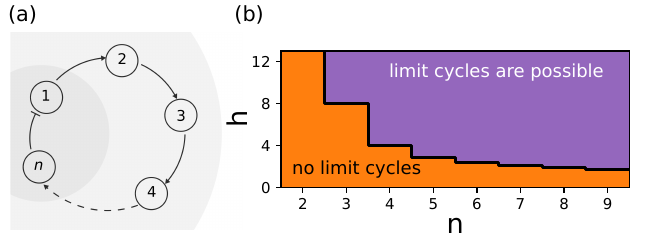}
    \caption{(a) Scheme of the model proposed by Goodwin. A biochemical negative feedback loop where each component is produced by a first order reaction from the previous component, with exception of the first one, whose zero-order reaction is repressed by the last component. (b) Secant condition. Minimal cooperativity that leads to oscillations for each length $n$ of the biochemical chain.}
    \label{figure_intro}
\end{figure}
The real world, however, is neither compartmental nor well-mixed. The cells are of small but finite size and inside them a great deal of inhomogeneity is present --localization of species occur within nanodomains \cite{fonkeu2019mrna,hafner2019localized,sun2021prevalence,wu2016translation,jackson2020camp}. 
While mRNA translation occurs in some place in the cytoplasm relatively distant to from the nucleus, it is in this nucleus where the gene generates the mRNA and where the resulting protein may exert its repressive influence. 
Therefore, one trip of a new born mRNA molecule to the cytoplasm sites populated by ribosomes and an another trip back to the nucleus for the resulting proteins are needed for the cycle of expression and repression to be completed. 
In the compartmental approach, these trips are introduced in the form of fictitious intermediate reaction steps: mRNA$_{Nucleus}$ $ \rightarrow$ mRNA$_{Cytoplasm}$, P$_{Cytoplasm}$ $ \rightarrow$ P$_{Nucleus}$, etc. each one providing an equation of a similar structure as Eq.(\ref{int_step}). 
A related and less realistic approach consists in introducing a prescribed delay in the equations\cite{novak2008,bordyugov2013}, with the purpose of mimicking molecular transport times as well as extra steps. 

In our work, in contrast, we investigate the impact of including a realistic model for the transport across the inhomogeneties of the cell based on the process of molecular diffusion and the result is surprising: a loop based on Eqs.(\ref{one-step}) does now oscillate. Not only does it, but it also requires a cooperativity $h$ much smaller than the simplest compartmental model capable or generating cellular rhythms.

In fact, the investigation of the role of heterogeneity in the emergence of oscillations in reaction diffusion systems whose homogeneous counterpart shows only intrinsic homeostatic equilibrium has a long history. For instance the case of pancreatic $\beta$-cells has been studied in \cite{Julyan_hetero}. More recently several cases of gene regulatory networks have been studied under the same light \cite{krause2018heterogeneity,naqib2012tunable,chaplain2015hopf,macnamara2016diffusion,macnamara2017spatio} 

In our model, as in real cells, the transcription and translation processes are located in spatially different regions. The gene (GEN) produces mRNA at the nucleus, well separated from the ribosomes (RIB), at which protein synthesis takes place. Now, both  mRNA and P migrate through diffusion, so that now the corresponding concentrations $m(\vec{r},t)$ and  $p(\vec{r},t)$  depend on space and time. Thus, Eq. (1) must be rewritten in order to account for diffusion, yielding 
\begin{eqnarray}
\label{activator-repressor-diffusion}
\dot{m}(\vec{r},t)   & = & \frac{\omega_m}{1+p(\vec{r},t)^h} \, f_{GEN}(\vec{r}) -\gamma_m \, m(\vec{r},t)  \nonumber \\
&  & + D_m \, \nabla^2 m(\vec{r},t)    \\
\dot{p}(\vec{r},t) & =& \omega_p \; f_{RIB}(\vec{r})  \, m(\vec{r},t) - \gamma_p \, p(\vec{r},t)  \nonumber \\
&  & + D_p \nabla^2 \, p(\vec{r},t) ,   \nonumber 
\end{eqnarray}
where $f_{GEN}(\vec{r})$ and $f_{RIB}(\vec{r})$ refer to the spatial distributions of the gene and the ribosomes, respectively.  $D_m$ and $D_p$ are the effective diffusion coefficients experienced by the mRNA and P molecules. The parameters $\omega_m$ and $\omega_P$ refer to the rates of production of mRNA and P. Eqs. (\ref{activator-repressor-diffusion}) reduce to the Goodwin model for two components if one neglects the dependence of the concentrations on the spatial coordinates (well-mixed case).

\begin{figure} 
    \includegraphics[width=3.3in]{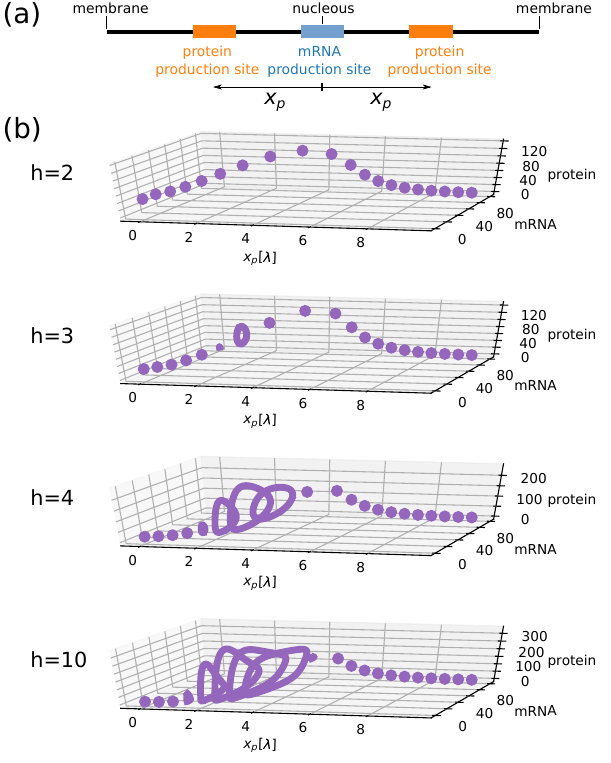}
    \caption{A modified version of the Goodwin model that accounts for diffusion of the components over a heterogeneous spatial domain, show oscillations that are dependent on the distance between the reaction zones. (a) Scheme of the model. (b) Trajectories after initial transient of the quantities $M(t)$ --as \emph{mRNA}-- and $P(t)$ --as \emph{protein}-- (s. Eq. \ref{integral_values}) for different distances and cooperativities. The higher the cooperativity $h$, the wider the range of distances where oscillations occur. The case $h=3$ shows oscillations for a narrow range of distances.  $ D = 0.1$ , $ \gamma = 0.1 $, $ R_{m} = R_{p} = 0.5$, $ R_{cell} = 10$ , $\omega_m = 10 $ , $\omega_p = 20 $   }
    \label{1d-diffusion}
\end{figure}

Of course, Eqs. \ref{activator-repressor-diffusion} can be generalized to include also intermediate reactions, but for the moment let us concentrate in the simpler case equivalent to the non oscillatory $n=2$ case of the Goodwin family \cite{goodwin65}. Let us also consider the dynamical behavior of the spatially one-dimensional version of  Eqs.(\ref{activator-repressor-diffusion}) obtained after the replacements $\vec{r}\rightarrow x$ and $\nabla^2\rightarrow \partial_{xx}$. We choose also the corresponding distributions $f_{GEN}(x)$ and $f_{RIB}(x)$ to be boxcar functions centered around the positions $x_m = 0$ and $x_p \neq 0$, with corresponding extents $2R_{m}$ and $2R_{p}$, respectively (see model details in the Supplemental Material).

For the sake of comparison with the compartmental models we compute the values of the total amount of $m$ and $p$ within the cell to be, a quantity that should be somehow proportional to the homonyms in the Goodwin's family: 
\begin{eqnarray}
\label{integral_values}
	M(t) & = & \int_\Omega m(x,t) dx \nonumber \\
	P(t) & = & \int_\Omega p(x,t) dx, 
\end{eqnarray}
where $\Omega$ is the cell volume.

In Fig.\ref{1d-diffusion} we plot the final fate of the evolution of $M(t)$ and $P(t)$ as $t\rightarrow\infty$ displayed for different values of $x_p$ and $h$. Remarkably, for intermediate values of distances between the sources, the system display sustained oscillations. However, for lower and higher distances, the oscillations disappear and the system converges to a steady state. Notice that the higher the cooperativity, the shorter the necessary distance between nucleus and ribosomes to produce oscillations, but even with values as small as $h=3$ a single two-component expression-repression reaction generates limit cycles in a relatively wide range of nucleus-ribosome distances. 
The outcome is already interesting because it underlines the fact that purely diffusive transport is enough to allow for oscillatory behavior in a negatively feedback control circuit of only two species. 
This fact is reminiscent of the instabilities found in reaction-diffusion spatial patterns, where homogeneous profiles are turned unstable under certain values of diffusion coefficients \cite{turing1952,maini2012turing}. 
But from the point of view of the application to cellular rhythms, the result unveils another surprise: the relatively small cooperativity parameter necessary to achieve oscillations.

\begin{figure}[b]
    \includegraphics[width=3.3in]{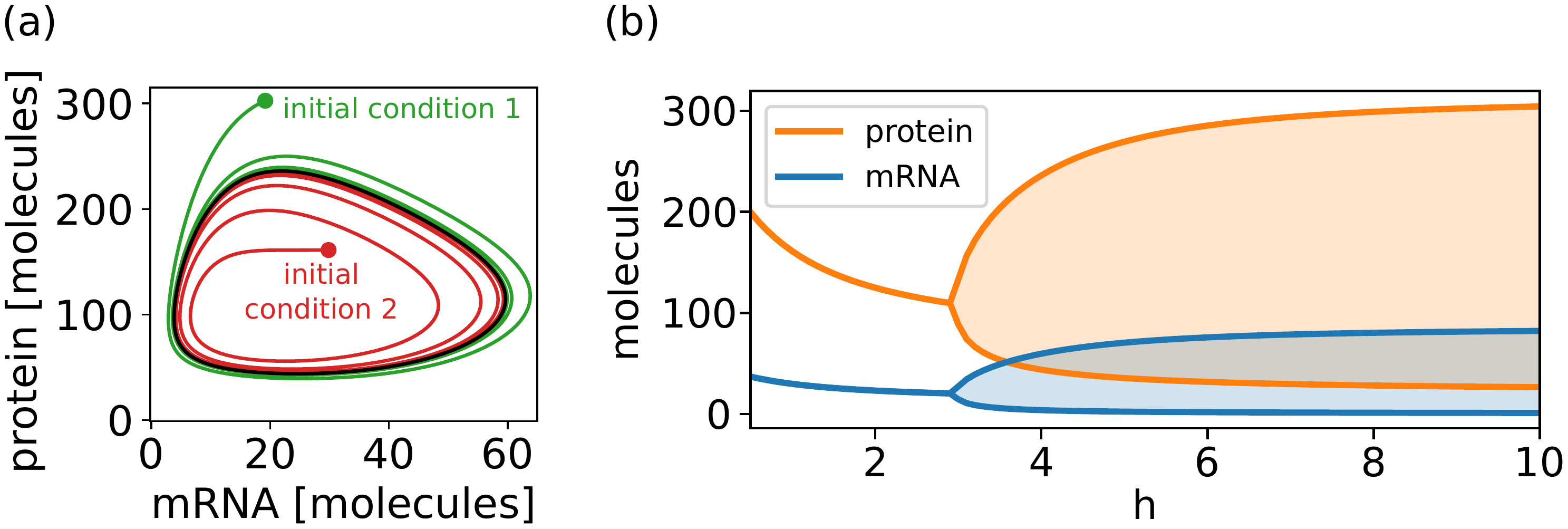}
    \caption{ (a) Parametric plot in terms of the total amounts of mRNA and protein ($M$ and $P$) for a single one dimensional transcription-translation feedback loop with two spatially separated reactions and diffusion, showing two example trajectories (red, green)converging to the limit cycle (black) for $h=4$. (b) Plot of minima and maxima (lines) as function of $h$, with shaded regions indicating oscillations. The onset of oscillations occur below $h=3$, a value achieved only in longer chains in the well-mixed case (Fig. \ref{figure_intro}). Parameters as in Fig. \ref{1d-diffusion}.} 
    \label{limit_cicle}
\end{figure}

In Fig.\ref{limit_cicle}a the temporal evolution of $M(t)$ and $P(t)$ corresponding to two different initial configurations of $m(x,t)$ and $p(x,t)$ is shown. 
The existence of a limit cycle is evident, a fact that confirms, on the one hand, the existence of regular stable oscillations in the system and on the other hand, a certain type of correspondence with a compartmental model of a number of variables at least larger than 3. 
The initial configurations in the considered cases are chosen to be $m(x,0) = m_0 $ and $ p(x,0) = p_0$ such that one configuration falls inside and the other outside the limit cycle. 
Fig.\ref{limit_cicle}b shows, for the same set of parameters corresponding to Fig.\ref{limit_cicle}a, the onset of oscillations as the cooperativity is increased from very low values, where no limit cycles are observed, to higher values, passing through a bifurcation as in the compartmental model. 
Recalling Fig.\ref{figure_intro}, in the compartmental model we need to take into account six components ($n=6$) to support oscillations with cooperativity $h=3$.

The onset of the oscillations is aided by the delays induced by the diffusive transport of the species. We derived an estimation of the time-scale associated with the diffusion process, by solving analytically a simplified problem, in which a punctual source of a single species is introduced as a perturbation at finite time in an infinite domain (see Supplemental Material). We termed it as relaxation time $t_R = (1/4\gamma) (\sqrt{1+4(x/\lambda)^2} -1)$,  which converges asymptotically to a linear dependence on the distance (related derivations can be found in \cite{berezhkovskii2011formation} and  \cite{ellery2012critical}, which named their results as local accumulation time and critical time, respectively). It is simple to verify this linear dependence in the spatio-temporal distribution of $m(x,t)$ and $p(x,t)$(see Supplemental Material).

\begin{figure*} 
    \includegraphics[width=6.6in]{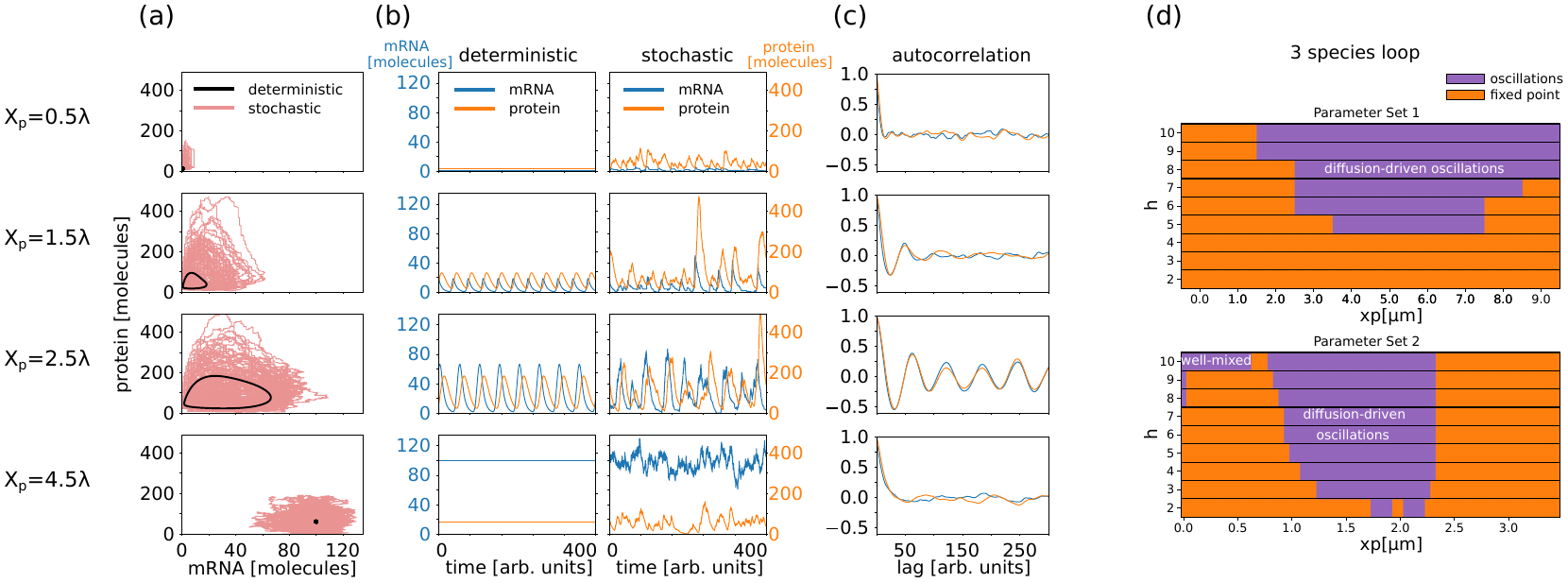}
    \caption{Stochastic (agent-based) and deterministic (partial differential equations) models show onset of oscillations of the number of $mRNA$ and $protein$ molecules in approximately the same range of separations. (a) Super-imposed trajectories from the deterministic and stochastic models. Limit cycles from the deterministic model and oscillatory noisy trajectories from the stochastic agent-based model appear in intermediate ranges of separation. (b) Time series of $mRNA$ and $ protein $ in a segment of the same trajectories. (c) The stochastic model displays noisy uncorrelated fluctuations for small and large distances between sources, as reflected in the autocorrelation function. For intermediate distances, the autocorrelation function clearly indicates oscillations, reflected in the peaks appearing at periodic lags . $h= 10 $, $ D = 0.1$ , $ \gamma = 0.1 $, $ R_{m} = R_{p} = 0.5$, $ R_{cell} = 10$ , $\omega_m = \omega_p = 10 $. (d) Example biochemical loops of three species show enhanced oscillations when sources of molecular species are separated. (upper panel) A fictitious feedback loop of three species constructed with a set of parameters compatible with experiments (s. Refs \cite{fonkeu2019mrna,sturrock2011spatio}) that would not oscillate in the well-mixed case, display oscillations if the distance between sources is appropriate. (lower panel) A feedback loop constructed with a set of parameters that yields oscillations in the compartmental models (well-mixed case) (s. Ref. \cite{gonze2013goodwin}), looses and regains oscillatory behavior when the distance between sources is increased, yielding a distinct diffusion-dominated oscillation parameter region. For low $h$, oscillations are only possible in the diffusion-dominated region .}
    \label{panel}
\end{figure*}

In the following we show that the central idea of our model -- the fact that self-sustained oscillations in cells can arise from the spatial separation between transcription and translation sites and without resorting to complex loops of intermediate reactions -- is not an artifact of the partial differential equations (PDE) representation of this scenario. An alternative simulation of this process can be achieved by modeling the individual trajectories of the reacting molecules by an stochastic Langevin equation\cite{erban2020}. This approach should include also a proper description  of the probability of reaction which we implement through an agent based algorithm. The deterministic results of the PDE should then be compared with the long term ensemble average of the stochastic model. 
We confirmed the existence of self-sustained oscillations by running simulations of the stochastic counterpart of the deterministic representation. In order to perform the simulations, we coupled reaction and diffusion of individual molecules through a spatial domain, by means of the Gillespie algorithm and the Langevin equation. Molecules are subjected to the set of reactions 

\begin{eqnarray}
\label{gillespie-react-diff}
mRNA     & \rightarrow   & \phi        \nonumber  \\
protein  & \rightarrow   & \phi         \nonumber \\
\phi     & \rightarrow  & mRNA           \\
\phi     & \rightarrow    & protein     \nonumber \\
x_i(t+\Delta t) & =&  x_i(t ) + \sqrt{2D \Delta t} \xi , \nonumber 
\end{eqnarray}
whereby $\phi$ denotes chemical species that are of no interest in order to represent degradation and creation of $mRNA$ and $protein$ molecules, $ x_i(t )$ is the position of the molecule $i$ at time $t$, $ \xi $ is a Gaussian white noise \cite{erban2020}. The reaction rates in each reaction are such that the system is equivalent to Eq.\ref{activator-repressor-diffusion} (see details in Supplemental Material). D is the diffusion coefficient.
In Fig.\ref{panel} we show the orbits resulting from both the deterministic and the stochastic approaches. Both stochastic and deterministic numerical results clearly show oscillations only for intermediate values of separation between sources, as it is evident from time series in both representations. In the stochastic case, the periodicity is sharply revealed by the peaks in the autocorrelation function of individual time series. Notice that the first peak is shifted to higher periods as the separation becomes greater. As in the case of the deterministic approach, this shift can be explained by the increased delay in the transport of molecules. 

Finally, we consider a loop that includes three molecular species, as in Eqs. \ref{goodwin_three_steps}. We choose two sets of parameters. In the first set, both kinetic and transport parameters are compatible with experimentally obtained measurements on biological systems \cite{fonkeu2019mrna,sturrock2011spatio}. In the second set, kinetic parameters are taken from published models that display oscillations in the well-mixed case \cite{gonze2013goodwin}(see complete parameter list in Supplemental Material). 
The production site for the third species is chosen to be identical to that for the second species, i.e. $x_r=x_p$, and the distance is defined to the production site of the first species. We identified the combination of $h$ and $x_p$ that yield oscillatory solutions. 
The results for the first set (Fig.\ref{panel}d--upper panel) demonstrate that a simple negative feedback loop in a biological cell can transition from homeostatic to oscillatory behavior for distances compatible with the dimensions of a cell, provided the cooperativity is at least 5. 
The results for the second set (Fig.\ref{panel}d--lower panel) reveal that a system that is tuned to display oscillations if there is no separation, can loose this feature if the distance between sources is increased, and recover it again if this separation is further increased. Furthermore, the oscillations appear in an extended region of parameters, where far smaller values of $h$ are required and amplitudes can be higher (see Supplemental Material). 
The disconnection between the two oscillatory regions of parameters, as well as their shape, implies that diffusion not only arises as a distinct mechanism to enable oscillations, but also improves their robustness.

Summarizing, we showed that a simple biochemical feedback loop, that in well-mixed conditions would be doomed to display homeostatic behavior, can be turned into an oscillator by physically separating the domains where reactions occur. This study can reconcile the parameter choice in mathematical modeling with the biologically plausible values. It can provide the modeling framework to recent experimental observations that suggest a crucial role of the localization of synthesis within cells \cite{fonkeu2019mrna,hafner2019localized,sun2021prevalence,wu2016translation,weitz2014diversity}.

PR is grateful to Claudia R. Arbeitman for useful discussions. M.E.G acknowledges support from the DFG through the Grant RTG 2749/1 "Multiscale Clocks".  O.P. acknowledges COST Action CA21169. Part of the present research was carried out within the framework of the activities of the Spanish Government through the "Maria de Maeztu Centre of Excellence accreditation to IMEDEA (CSIC-UIB) (CEX2021-001198). MEG and PR acknowledge support from the  University of Kassel (ZFF-PROJEKT 2377) and PR from the Joachim Herz Stiftung. 
\bibliography{diff_clocks}

\providecommand{\noopsort}[1]{}\providecommand{\singleletter}[1]{#1}%
\begin{thebibliography}{28}%
\makeatletter
\providecommand \@ifxundefined [1]{%
 \@ifx{#1\undefined}
}%
\providecommand \@ifnum [1]{%
 \ifnum #1\expandafter \@firstoftwo
 \else \expandafter \@secondoftwo
 \fi
}%
\providecommand \@ifx [1]{%
 \ifx #1\expandafter \@firstoftwo
 \else \expandafter \@secondoftwo
 \fi
}%
\providecommand \natexlab [1]{#1}%
\providecommand \enquote  [1]{``#1''}%
\providecommand \bibnamefont  [1]{#1}%
\providecommand \bibfnamefont [1]{#1}%
\providecommand \citenamefont [1]{#1}%
\providecommand \href@noop [0]{\@secondoftwo}%
\providecommand \href [0]{\begingroup \@sanitize@url \@href}%
\providecommand \@href[1]{\@@startlink{#1}\@@href}%
\providecommand \@@href[1]{\endgroup#1\@@endlink}%
\providecommand \@sanitize@url [0]{\catcode `\\12\catcode `\$12\catcode
  `\&12\catcode `\#12\catcode `\^12\catcode `\_12\catcode `\%12\relax}%
\providecommand \@@startlink[1]{}%
\providecommand \@@endlink[0]{}%
\providecommand \url  [0]{\begingroup\@sanitize@url \@url }%
\providecommand \@url [1]{\endgroup\@href {#1}{\urlprefix }}%
\providecommand \urlprefix  [0]{URL }%
\providecommand \Eprint [0]{\href }%
\providecommand \doibase [0]{https://doi.org/}%
\providecommand \selectlanguage [0]{\@gobble}%
\providecommand \bibinfo  [0]{\@secondoftwo}%
\providecommand \bibfield  [0]{\@secondoftwo}%
\providecommand \translation [1]{[#1]}%
\providecommand \BibitemOpen [0]{}%
\providecommand \bibitemStop [0]{}%
\providecommand \bibitemNoStop [0]{.\EOS\space}%
\providecommand \EOS [0]{\spacefactor3000\relax}%
\providecommand \BibitemShut  [1]{\csname bibitem#1\endcsname}%
\let\auto@bib@innerbib\@empty
\bibitem [{\citenamefont {Forger}(2017)}]{forger2017biological}%
  \BibitemOpen
  \bibfield  {author} {\bibinfo {author} {\bibfnamefont {D.~B.}\ \bibnamefont
  {Forger}},\ }\href@noop {} {\emph {\bibinfo {title} {Biological Clocks,
  Rhythms, and Oscillations: The Theory of Biological Timekeeping}}}\ (\bibinfo
   {publisher} {MIT Press},\ \bibinfo {year} {2017})\BibitemShut {NoStop}%
\bibitem [{\citenamefont {Gonze}\ \emph {et~al.}(2002)\citenamefont {Gonze},
  \citenamefont {Halloy},\ and\ \citenamefont {Goldbeter}}]{gonze2002}%
  \BibitemOpen
  \bibfield  {author} {\bibinfo {author} {\bibfnamefont {D.}~\bibnamefont
  {Gonze}}, \bibinfo {author} {\bibfnamefont {J.}~\bibnamefont {Halloy}},\ and\
  \bibinfo {author} {\bibfnamefont {A.}~\bibnamefont {Goldbeter}},\ }\bibfield
  {title} {\bibinfo {title} {Deterministic versus stochastic models for
  circadian rhythms},\ }\href@noop {} {\bibfield  {journal} {\bibinfo
  {journal} {Journal of Biological Physics}\ }\textbf {\bibinfo {volume}
  {28}},\ \bibinfo {pages} {637} (\bibinfo {year} {2002})}\BibitemShut
  {NoStop}%
\bibitem [{\citenamefont {Nov{\'a}k}\ and\ \citenamefont
  {Tyson}(2008)}]{novak2008}%
  \BibitemOpen
  \bibfield  {author} {\bibinfo {author} {\bibfnamefont {B.}~\bibnamefont
  {Nov{\'a}k}}\ and\ \bibinfo {author} {\bibfnamefont {J.~J.}\ \bibnamefont
  {Tyson}},\ }\bibfield  {title} {\bibinfo {title} {Design principles of
  biochemical oscillators},\ }\href@noop {} {\bibfield  {journal} {\bibinfo
  {journal} {Nature Reviews Molecular Cell Biology}\ }\textbf {\bibinfo
  {volume} {9}},\ \bibinfo {pages} {981} (\bibinfo {year} {2008})}\BibitemShut
  {NoStop}%
\bibitem [{\citenamefont {Ananthasubramaniam}\ and\ \citenamefont
  {Herzel}(2014)}]{bharath2014}%
  \BibitemOpen
  \bibfield  {author} {\bibinfo {author} {\bibfnamefont {B.}~\bibnamefont
  {Ananthasubramaniam}}\ and\ \bibinfo {author} {\bibfnamefont
  {H.}~\bibnamefont {Herzel}},\ }\bibfield  {title} {\bibinfo {title} {Positive
  feedback promotes oscillations in negative feedback loops},\ }\href@noop {}
  {\bibfield  {journal} {\bibinfo  {journal} {PLoS One}\ }\textbf {\bibinfo
  {volume} {9}},\ \bibinfo {pages} {e104761} (\bibinfo {year}
  {2014})}\BibitemShut {NoStop}%
\bibitem [{\citenamefont {Goldbeter}(1995)}]{goldbeter1995model}%
  \BibitemOpen
  \bibfield  {author} {\bibinfo {author} {\bibfnamefont {A.}~\bibnamefont
  {Goldbeter}},\ }\bibfield  {title} {\bibinfo {title} {A model for circadian
  oscillations in the {Drosophila} period protein ({PER})},\ }\href@noop {}
  {\bibfield  {journal} {\bibinfo  {journal} {Proceedings of the Royal Society
  of London. Series B: Biological Sciences}\ }\textbf {\bibinfo {volume}
  {261}},\ \bibinfo {pages} {319} (\bibinfo {year} {1995})}\BibitemShut
  {NoStop}%
\bibitem [{\citenamefont {Griffith}(1968)}]{griffith68}%
  \BibitemOpen
  \bibfield  {author} {\bibinfo {author} {\bibfnamefont {J.~S.}\ \bibnamefont
  {Griffith}},\ }\bibfield  {title} {\bibinfo {title} {Mathematics of cellular
  control processes {I}. negative feedback to one gene},\ }\href@noop {}
  {\bibfield  {journal} {\bibinfo  {journal} {Journal of Theoretical Biology}\
  }\textbf {\bibinfo {volume} {20}},\ \bibinfo {pages} {202} (\bibinfo {year}
  {1968})}\BibitemShut {NoStop}%
\bibitem [{\citenamefont {Goodwin}(1965)}]{goodwin65}%
  \BibitemOpen
  \bibfield  {author} {\bibinfo {author} {\bibfnamefont {B.~C.}\ \bibnamefont
  {Goodwin}},\ }\bibfield  {title} {\bibinfo {title} {Oscillatory behavior in
  enzymatic control processes},\ }\href@noop {} {\bibfield  {journal} {\bibinfo
   {journal} {Advances in Enzyme Regulation}\ }\textbf {\bibinfo {volume}
  {3}},\ \bibinfo {pages} {425} (\bibinfo {year} {1965})}\BibitemShut {NoStop}%
\bibitem [{\citenamefont {Tyson}\ and\ \citenamefont
  {Othmer}(1978)}]{tyson1978dynamics}%
  \BibitemOpen
  \bibfield  {author} {\bibinfo {author} {\bibfnamefont {J.~J.}\ \bibnamefont
  {Tyson}}\ and\ \bibinfo {author} {\bibfnamefont {H.~G.}\ \bibnamefont
  {Othmer}},\ }\bibfield  {title} {\bibinfo {title} {The dynamics of feedback
  control circuits in biochemical pathways},\ }\href@noop {} {\bibfield
  {journal} {\bibinfo  {journal} {Progress in theoretical biology}\ }\textbf
  {\bibinfo {volume} {5}},\ \bibinfo {pages} {1} (\bibinfo {year}
  {1978})}\BibitemShut {NoStop}%
\bibitem [{\citenamefont {Fonkeu}\ \emph {et~al.}(2019)\citenamefont {Fonkeu},
  \citenamefont {Kraynyukova}, \citenamefont {Hafner}, \citenamefont {Kochen},
  \citenamefont {Sartori}, \citenamefont {Schuman},\ and\ \citenamefont
  {Tchumatchenko}}]{fonkeu2019mrna}%
  \BibitemOpen
  \bibfield  {author} {\bibinfo {author} {\bibfnamefont {Y.}~\bibnamefont
  {Fonkeu}}, \bibinfo {author} {\bibfnamefont {N.}~\bibnamefont {Kraynyukova}},
  \bibinfo {author} {\bibfnamefont {A.-S.}\ \bibnamefont {Hafner}}, \bibinfo
  {author} {\bibfnamefont {L.}~\bibnamefont {Kochen}}, \bibinfo {author}
  {\bibfnamefont {F.}~\bibnamefont {Sartori}}, \bibinfo {author} {\bibfnamefont
  {E.~M.}\ \bibnamefont {Schuman}},\ and\ \bibinfo {author} {\bibfnamefont
  {T.}~\bibnamefont {Tchumatchenko}},\ }\bibfield  {title} {\bibinfo {title}
  {How {mRNA} localization and protein synthesis sites influence dendritic
  protein distribution and dynamics},\ }\href@noop {} {\bibfield  {journal}
  {\bibinfo  {journal} {Neuron}\ }\textbf {\bibinfo {volume} {103}},\ \bibinfo
  {pages} {1109} (\bibinfo {year} {2019})}\BibitemShut {NoStop}%
\bibitem [{\citenamefont {Hafner}\ \emph {et~al.}(2019)\citenamefont {Hafner},
  \citenamefont {Donlin-Asp}, \citenamefont {Leitch}, \citenamefont {Herzog},\
  and\ \citenamefont {Schuman}}]{hafner2019localized}%
  \BibitemOpen
  \bibfield  {author} {\bibinfo {author} {\bibfnamefont {A.-S.}\ \bibnamefont
  {Hafner}}, \bibinfo {author} {\bibfnamefont {P.~G.}\ \bibnamefont
  {Donlin-Asp}}, \bibinfo {author} {\bibfnamefont {B.}~\bibnamefont {Leitch}},
  \bibinfo {author} {\bibfnamefont {E.}~\bibnamefont {Herzog}},\ and\ \bibinfo
  {author} {\bibfnamefont {E.~M.}\ \bibnamefont {Schuman}},\ }\bibfield
  {title} {\bibinfo {title} {Local protein synthesis is a ubiquitous feature of
  neuronal pre- and postsynaptic compartments},\ }\href@noop {} {\bibfield
  {journal} {\bibinfo  {journal} {Science}\ }\textbf {\bibinfo {volume}
  {364}},\ \bibinfo {pages} {eaau3644} (\bibinfo {year} {2019})}\BibitemShut
  {NoStop}%
\bibitem [{\citenamefont {Sun}\ \emph {et~al.}(2021)\citenamefont {Sun},
  \citenamefont {Nold}, \citenamefont {Fusco}, \citenamefont {Rangaraju},
  \citenamefont {Tchumatchenko}, \citenamefont {Heilemann},\ and\ \citenamefont
  {Schuman}}]{sun2021prevalence}%
  \BibitemOpen
  \bibfield  {author} {\bibinfo {author} {\bibfnamefont {C.}~\bibnamefont
  {Sun}}, \bibinfo {author} {\bibfnamefont {A.}~\bibnamefont {Nold}}, \bibinfo
  {author} {\bibfnamefont {C.~M.}\ \bibnamefont {Fusco}}, \bibinfo {author}
  {\bibfnamefont {V.}~\bibnamefont {Rangaraju}}, \bibinfo {author}
  {\bibfnamefont {T.}~\bibnamefont {Tchumatchenko}}, \bibinfo {author}
  {\bibfnamefont {M.}~\bibnamefont {Heilemann}},\ and\ \bibinfo {author}
  {\bibfnamefont {E.~M.}\ \bibnamefont {Schuman}},\ }\bibfield  {title}
  {\bibinfo {title} {The prevalence and specificity of local protein synthesis
  during neuronal synaptic plasticity},\ }\href@noop {} {\bibfield  {journal}
  {\bibinfo  {journal} {Science Advances}\ }\textbf {\bibinfo {volume} {7}},\
  \bibinfo {pages} {eabj0790} (\bibinfo {year} {2021})}\BibitemShut {NoStop}%
\bibitem [{\citenamefont {Wu}\ \emph {et~al.}(2016)\citenamefont {Wu},
  \citenamefont {Eliscovich}, \citenamefont {Yoon},\ and\ \citenamefont
  {Singer}}]{wu2016translation}%
  \BibitemOpen
  \bibfield  {author} {\bibinfo {author} {\bibfnamefont {B.}~\bibnamefont
  {Wu}}, \bibinfo {author} {\bibfnamefont {C.}~\bibnamefont {Eliscovich}},
  \bibinfo {author} {\bibfnamefont {Y.~J.}\ \bibnamefont {Yoon}},\ and\
  \bibinfo {author} {\bibfnamefont {R.~H.}\ \bibnamefont {Singer}},\ }\bibfield
   {title} {\bibinfo {title} {Translation dynamics of single {mRNAs} in live
  cells and neurons},\ }\href@noop {} {\bibfield  {journal} {\bibinfo
  {journal} {Science}\ }\textbf {\bibinfo {volume} {352}},\ \bibinfo {pages}
  {1430} (\bibinfo {year} {2016})}\BibitemShut {NoStop}%
\bibitem [{\citenamefont {Jackson}(2020)}]{jackson2020camp}%
  \BibitemOpen
  \bibfield  {author} {\bibinfo {author} {\bibfnamefont {P.~K.}\ \bibnamefont
  {Jackson}},\ }\bibfield  {title} {\bibinfo {title} {{cAMP} signaling in
  nanodomains},\ }\href@noop {} {\bibfield  {journal} {\bibinfo  {journal}
  {Cell}\ }\textbf {\bibinfo {volume} {182}},\ \bibinfo {pages} {1379}
  (\bibinfo {year} {2020})}\BibitemShut {NoStop}%
\bibitem [{\citenamefont {Bordyugov}\ \emph {et~al.}(2013)\citenamefont
  {Bordyugov}, \citenamefont {Westermark}, \citenamefont
  {Koren{\v{c}}i{\v{c}}}, \citenamefont {Bernard},\ and\ \citenamefont
  {Herzel}}]{bordyugov2013}%
  \BibitemOpen
  \bibfield  {author} {\bibinfo {author} {\bibfnamefont {G.}~\bibnamefont
  {Bordyugov}}, \bibinfo {author} {\bibfnamefont {P.~O.}\ \bibnamefont
  {Westermark}}, \bibinfo {author} {\bibfnamefont {A.}~\bibnamefont
  {Koren{\v{c}}i{\v{c}}}}, \bibinfo {author} {\bibfnamefont {S.}~\bibnamefont
  {Bernard}},\ and\ \bibinfo {author} {\bibfnamefont {H.}~\bibnamefont
  {Herzel}},\ }\bibinfo {title} {Mathematical modeling in chronobiology},\ in\
  \href {https://doi.org/10.1007/978-3-642-25950-0_14} {\emph {\bibinfo
  {booktitle} {Circadian Clocks}}},\ \bibinfo {editor} {edited by\ \bibinfo
  {editor} {\bibfnamefont {A.}~\bibnamefont {Kramer}}\ and\ \bibinfo {editor}
  {\bibfnamefont {M.}~\bibnamefont {Merrow}}}\ (\bibinfo  {publisher}
  {Springer},\ \bibinfo {year} {2013})\ pp.\ \bibinfo {pages}
  {335--357}\BibitemShut {NoStop}%
\bibitem [{\citenamefont {Cartwright}(2000)}]{Julyan_hetero}%
  \BibitemOpen
  \bibfield  {author} {\bibinfo {author} {\bibfnamefont {J.~H.~E.}\
  \bibnamefont {Cartwright}},\ }\bibfield  {title} {\bibinfo {title} {Emergent
  global oscillations in heterogeneous excitable media: The example of
  pancreatic $\ensuremath{\beta}$ cells},\ }\href
  {https://doi.org/10.1103/PhysRevE.62.1149} {\bibfield  {journal} {\bibinfo
  {journal} {Phys. Rev. E}\ }\textbf {\bibinfo {volume} {62}},\ \bibinfo
  {pages} {1149} (\bibinfo {year} {2000})}\BibitemShut {NoStop}%
\bibitem [{\citenamefont {Krause}\ \emph {et~al.}(2018)\citenamefont {Krause},
  \citenamefont {Klika}, \citenamefont {Woolley},\ and\ \citenamefont
  {Gaffney}}]{krause2018heterogeneity}%
  \BibitemOpen
  \bibfield  {author} {\bibinfo {author} {\bibfnamefont {A.~L.}\ \bibnamefont
  {Krause}}, \bibinfo {author} {\bibfnamefont {V.}~\bibnamefont {Klika}},
  \bibinfo {author} {\bibfnamefont {T.~E.}\ \bibnamefont {Woolley}},\ and\
  \bibinfo {author} {\bibfnamefont {E.~A.}\ \bibnamefont {Gaffney}},\
  }\bibfield  {title} {\bibinfo {title} {Heterogeneity induces spatiotemporal
  oscillations in reaction-diffusion systems},\ }\href@noop {} {\bibfield
  {journal} {\bibinfo  {journal} {Physical Review E}\ }\textbf {\bibinfo
  {volume} {97}},\ \bibinfo {pages} {052206} (\bibinfo {year}
  {2018})}\BibitemShut {NoStop}%
\bibitem [{\citenamefont {Naqib}\ \emph {et~al.}(2012)\citenamefont {Naqib},
  \citenamefont {Quail}, \citenamefont {Musa}, \citenamefont {Vulpe},
  \citenamefont {Nadeau}, \citenamefont {Lei},\ and\ \citenamefont
  {Glass}}]{naqib2012tunable}%
  \BibitemOpen
  \bibfield  {author} {\bibinfo {author} {\bibfnamefont {F.}~\bibnamefont
  {Naqib}}, \bibinfo {author} {\bibfnamefont {T.}~\bibnamefont {Quail}},
  \bibinfo {author} {\bibfnamefont {L.}~\bibnamefont {Musa}}, \bibinfo {author}
  {\bibfnamefont {H.}~\bibnamefont {Vulpe}}, \bibinfo {author} {\bibfnamefont
  {J.}~\bibnamefont {Nadeau}}, \bibinfo {author} {\bibfnamefont
  {J.}~\bibnamefont {Lei}},\ and\ \bibinfo {author} {\bibfnamefont
  {L.}~\bibnamefont {Glass}},\ }\bibfield  {title} {\bibinfo {title} {Tunable
  oscillations and chaotic dynamics in systems with localized synthesis},\
  }\href {https://doi.org/10.1103/PhysRevE.85.046210} {\bibfield  {journal}
  {\bibinfo  {journal} {Phys. Rev. E}\ }\textbf {\bibinfo {volume} {85}},\
  \bibinfo {pages} {046210} (\bibinfo {year} {2012})}\BibitemShut {NoStop}%
\bibitem [{\citenamefont {Chaplain}\ \emph {et~al.}(2015)\citenamefont
  {Chaplain}, \citenamefont {Ptashnyk},\ and\ \citenamefont
  {Sturrock}}]{chaplain2015hopf}%
  \BibitemOpen
  \bibfield  {author} {\bibinfo {author} {\bibfnamefont {M.}~\bibnamefont
  {Chaplain}}, \bibinfo {author} {\bibfnamefont {M.}~\bibnamefont {Ptashnyk}},\
  and\ \bibinfo {author} {\bibfnamefont {M.}~\bibnamefont {Sturrock}},\
  }\bibfield  {title} {\bibinfo {title} {Hopf bifurcation in a gene regulatory
  network model: {Molecular} movement causes oscillations},\ }\href@noop {}
  {\bibfield  {journal} {\bibinfo  {journal} {Mathematical Models and Methods
  in Applied Sciences}\ }\textbf {\bibinfo {volume} {25}},\ \bibinfo {pages}
  {1179} (\bibinfo {year} {2015})}\BibitemShut {NoStop}%
\bibitem [{\citenamefont {Macnamara}\ and\ \citenamefont
  {Chaplain}(2016)}]{macnamara2016diffusion}%
  \BibitemOpen
  \bibfield  {author} {\bibinfo {author} {\bibfnamefont {C.~K.}\ \bibnamefont
  {Macnamara}}\ and\ \bibinfo {author} {\bibfnamefont {M.~A.}\ \bibnamefont
  {Chaplain}},\ }\bibfield  {title} {\bibinfo {title} {Diffusion driven
  oscillations in gene regulatory networks},\ }\href@noop {} {\bibfield
  {journal} {\bibinfo  {journal} {Journal of Theoretical Biology}\ }\textbf
  {\bibinfo {volume} {407}},\ \bibinfo {pages} {51} (\bibinfo {year}
  {2016})}\BibitemShut {NoStop}%
\bibitem [{\citenamefont {Macnamara}\ and\ \citenamefont
  {Chaplain}(2017)}]{macnamara2017spatio}%
  \BibitemOpen
  \bibfield  {author} {\bibinfo {author} {\bibfnamefont {C.~K.}\ \bibnamefont
  {Macnamara}}\ and\ \bibinfo {author} {\bibfnamefont {M.~A.}\ \bibnamefont
  {Chaplain}},\ }\bibfield  {title} {\bibinfo {title} {Spatio-temporal models
  of synthetic genetic oscillators},\ }\href@noop {} {\bibfield  {journal}
  {\bibinfo  {journal} {Mathematical Biosciences and Engineering}\ }\textbf
  {\bibinfo {volume} {14}},\ \bibinfo {pages} {249} (\bibinfo {year}
  {2017})}\BibitemShut {NoStop}%
\bibitem [{\citenamefont {Turing}(1952)}]{turing1952}%
  \BibitemOpen
  \bibfield  {author} {\bibinfo {author} {\bibfnamefont {A.~M.}\ \bibnamefont
  {Turing}},\ }\bibfield  {title} {\bibinfo {title} {The chemical basis of
  morphogenesis},\ }\href@noop {} {\bibfield  {journal} {\bibinfo  {journal}
  {Philosophical Transactions of the Royal Society of London. Series B,
  Biological Sciences}\ }\textbf {\bibinfo {volume} {237}},\ \bibinfo {pages}
  {37} (\bibinfo {year} {1952})}\BibitemShut {NoStop}%
\bibitem [{\citenamefont {Maini}\ \emph {et~al.}(2012)\citenamefont {Maini},
  \citenamefont {Woolley}, \citenamefont {Baker}, \citenamefont {Gaffney},\
  and\ \citenamefont {Lee}}]{maini2012turing}%
  \BibitemOpen
  \bibfield  {author} {\bibinfo {author} {\bibfnamefont {P.~K.}\ \bibnamefont
  {Maini}}, \bibinfo {author} {\bibfnamefont {T.~E.}\ \bibnamefont {Woolley}},
  \bibinfo {author} {\bibfnamefont {R.~E.}\ \bibnamefont {Baker}}, \bibinfo
  {author} {\bibfnamefont {E.~A.}\ \bibnamefont {Gaffney}},\ and\ \bibinfo
  {author} {\bibfnamefont {S.~S.}\ \bibnamefont {Lee}},\ }\bibfield  {title}
  {\bibinfo {title} {Turing's model for biological pattern formation and the
  robustness problem},\ }\href@noop {} {\bibfield  {journal} {\bibinfo
  {journal} {Interface Focus}\ }\textbf {\bibinfo {volume} {2}},\ \bibinfo
  {pages} {487} (\bibinfo {year} {2012})}\BibitemShut {NoStop}%
\bibitem [{\citenamefont {Berezhkovskii}\ \emph {et~al.}(2011)\citenamefont
  {Berezhkovskii}, \citenamefont {Sample},\ and\ \citenamefont
  {Shvartsman}}]{berezhkovskii2011formation}%
  \BibitemOpen
  \bibfield  {author} {\bibinfo {author} {\bibfnamefont {A.~M.}\ \bibnamefont
  {Berezhkovskii}}, \bibinfo {author} {\bibfnamefont {C.}~\bibnamefont
  {Sample}},\ and\ \bibinfo {author} {\bibfnamefont {S.~Y.}\ \bibnamefont
  {Shvartsman}},\ }\bibfield  {title} {\bibinfo {title} {Formation of morphogen
  gradients: Local accumulation time},\ }\href@noop {} {\bibfield  {journal}
  {\bibinfo  {journal} {Physical Review E}\ }\textbf {\bibinfo {volume} {83}},\
  \bibinfo {pages} {051906} (\bibinfo {year} {2011})}\BibitemShut {NoStop}%
\bibitem [{\citenamefont {Ellery}\ \emph {et~al.}(2012)\citenamefont {Ellery},
  \citenamefont {Simpson}, \citenamefont {McCue},\ and\ \citenamefont
  {Baker}}]{ellery2012critical}%
  \BibitemOpen
  \bibfield  {author} {\bibinfo {author} {\bibfnamefont {A.~J.}\ \bibnamefont
  {Ellery}}, \bibinfo {author} {\bibfnamefont {M.~J.}\ \bibnamefont {Simpson}},
  \bibinfo {author} {\bibfnamefont {S.~W.}\ \bibnamefont {McCue}},\ and\
  \bibinfo {author} {\bibfnamefont {R.~E.}\ \bibnamefont {Baker}},\ }\bibfield
  {title} {\bibinfo {title} {Critical time scales for
  advection-diffusion-reaction processes},\ }\href@noop {} {\bibfield
  {journal} {\bibinfo  {journal} {Physical Review E}\ }\textbf {\bibinfo
  {volume} {85}},\ \bibinfo {pages} {041135} (\bibinfo {year}
  {2012})}\BibitemShut {NoStop}%
\bibitem [{\citenamefont {Erban}\ and\ \citenamefont
  {Chapman}(2020)}]{erban2020}%
  \BibitemOpen
  \bibfield  {author} {\bibinfo {author} {\bibfnamefont {R.}~\bibnamefont
  {Erban}}\ and\ \bibinfo {author} {\bibfnamefont {S.~J.}\ \bibnamefont
  {Chapman}},\ }\href@noop {} {\emph {\bibinfo {title} {Stochastic modelling of
  reaction--diffusion processes}}}\ (\bibinfo  {publisher} {Cambridge
  University Press},\ \bibinfo {year} {2020})\BibitemShut {NoStop}%
\bibitem [{\citenamefont {Sturrock}\ \emph {et~al.}(2011)\citenamefont
  {Sturrock}, \citenamefont {Terry}, \citenamefont {Xirodimas}, \citenamefont
  {Thompson},\ and\ \citenamefont {Chaplain}}]{sturrock2011spatio}%
  \BibitemOpen
  \bibfield  {author} {\bibinfo {author} {\bibfnamefont {M.}~\bibnamefont
  {Sturrock}}, \bibinfo {author} {\bibfnamefont {A.~J.}\ \bibnamefont {Terry}},
  \bibinfo {author} {\bibfnamefont {D.~P.}\ \bibnamefont {Xirodimas}}, \bibinfo
  {author} {\bibfnamefont {A.~M.}\ \bibnamefont {Thompson}},\ and\ \bibinfo
  {author} {\bibfnamefont {M.~A.}\ \bibnamefont {Chaplain}},\ }\bibfield
  {title} {\bibinfo {title} {Spatio-temporal modelling of the hes1 and p53-mdm2
  intracellular signalling pathways},\ }\href@noop {} {\bibfield  {journal}
  {\bibinfo  {journal} {Journal of Theoretical Biology}\ }\textbf {\bibinfo
  {volume} {273}},\ \bibinfo {pages} {15} (\bibinfo {year} {2011})}\BibitemShut
  {NoStop}%
\bibitem [{\citenamefont {Gonze}\ and\ \citenamefont
  {Abou-Jaoud{\'e}}(2013)}]{gonze2013goodwin}%
  \BibitemOpen
  \bibfield  {author} {\bibinfo {author} {\bibfnamefont {D.}~\bibnamefont
  {Gonze}}\ and\ \bibinfo {author} {\bibfnamefont {W.}~\bibnamefont
  {Abou-Jaoud{\'e}}},\ }\bibfield  {title} {\bibinfo {title} {The {G}oodwin
  model: behind the {H}ill function},\ }\href@noop {} {\bibfield  {journal}
  {\bibinfo  {journal} {PLoS one}\ }\textbf {\bibinfo {volume} {8}},\ \bibinfo
  {pages} {e69573} (\bibinfo {year} {2013})}\BibitemShut {NoStop}%
\bibitem [{\citenamefont {Weitz}\ \emph {et~al.}(2014)\citenamefont {Weitz},
  \citenamefont {Kim}, \citenamefont {Kapsner}, \citenamefont {Winfree},
  \citenamefont {Franco},\ and\ \citenamefont {Simmel}}]{weitz2014diversity}%
  \BibitemOpen
  \bibfield  {author} {\bibinfo {author} {\bibfnamefont {M.}~\bibnamefont
  {Weitz}}, \bibinfo {author} {\bibfnamefont {J.}~\bibnamefont {Kim}}, \bibinfo
  {author} {\bibfnamefont {K.}~\bibnamefont {Kapsner}}, \bibinfo {author}
  {\bibfnamefont {E.}~\bibnamefont {Winfree}}, \bibinfo {author} {\bibfnamefont
  {E.}~\bibnamefont {Franco}},\ and\ \bibinfo {author} {\bibfnamefont {F.~C.}\
  \bibnamefont {Simmel}},\ }\bibfield  {title} {\bibinfo {title} {Diversity in
  the dynamical behaviour of a compartmentalized programmable biochemical
  oscillator},\ }\href@noop {} {\bibfield  {journal} {\bibinfo  {journal}
  {Nature Chemistry}\ }\textbf {\bibinfo {volume} {6}},\ \bibinfo {pages} {295}
  (\bibinfo {year} {2014})}\BibitemShut {NoStop}%
\end{thebibliography}%


\begin{thebibliography}{7}%
\makeatletter
\providecommand \@ifxundefined [1]{%
 \@ifx{#1\undefined}
}%
\providecommand \@ifnum [1]{%
 \ifnum #1\expandafter \@firstoftwo
 \else \expandafter \@secondoftwo
 \fi
}%
\providecommand \@ifx [1]{%
 \ifx #1\expandafter \@firstoftwo
 \else \expandafter \@secondoftwo
 \fi
}%
\providecommand \natexlab [1]{#1}%
\providecommand \enquote  [1]{``#1''}%
\providecommand \bibnamefont  [1]{#1}%
\providecommand \bibfnamefont [1]{#1}%
\providecommand \citenamefont [1]{#1}%
\providecommand \href@noop [0]{\@secondoftwo}%
\providecommand \href [0]{\begingroup \@sanitize@url \@href}%
\providecommand \@href[1]{\@@startlink{#1}\@@href}%
\providecommand \@@href[1]{\endgroup#1\@@endlink}%
\providecommand \@sanitize@url [0]{\catcode `\\12\catcode `\$12\catcode
  `\&12\catcode `\#12\catcode `\^12\catcode `\_12\catcode `\%12\relax}%
\providecommand \@@startlink[1]{}%
\providecommand \@@endlink[0]{}%
\providecommand \url  [0]{\begingroup\@sanitize@url \@url }%
\providecommand \@url [1]{\endgroup\@href {#1}{\urlprefix }}%
\providecommand \urlprefix  [0]{URL }%
\providecommand \Eprint [0]{\href }%
\providecommand \doibase [0]{https://doi.org/}%
\providecommand \selectlanguage [0]{\@gobble}%
\providecommand \bibinfo  [0]{\@secondoftwo}%
\providecommand \bibfield  [0]{\@secondoftwo}%
\providecommand \translation [1]{[#1]}%
\providecommand \BibitemOpen [0]{}%
\providecommand \bibitemStop [0]{}%
\providecommand \bibitemNoStop [0]{.\EOS\space}%
\providecommand \EOS [0]{\spacefactor3000\relax}%
\providecommand \BibitemShut  [1]{\csname bibitem#1\endcsname}%
\let\auto@bib@innerbib\@empty
\bibitem [{\citenamefont {Goodwin}(1965)}]{goodwin65}%
  \BibitemOpen
  \bibfield  {author} {\bibinfo {author} {\bibfnamefont {B.~C.}\ \bibnamefont
  {Goodwin}},\ }\bibfield  {title} {\bibinfo {title} {Oscillatory behavior in
  enzymatic control processes},\ }\href@noop {} {\bibfield  {journal} {\bibinfo
   {journal} {Advances in Enzyme Regulation}\ }\textbf {\bibinfo {volume}
  {3}},\ \bibinfo {pages} {425} (\bibinfo {year} {1965})}\BibitemShut {NoStop}%
\bibitem [{\citenamefont {Griffith}(1968)}]{griffith68}%
  \BibitemOpen
  \bibfield  {author} {\bibinfo {author} {\bibfnamefont {J.~S.}\ \bibnamefont
  {Griffith}},\ }\bibfield  {title} {\bibinfo {title} {Mathematics of cellular
  control processes {I}. negative feedback to one gene},\ }\href@noop {}
  {\bibfield  {journal} {\bibinfo  {journal} {Journal of Theoretical Biology}\
  }\textbf {\bibinfo {volume} {20}},\ \bibinfo {pages} {202} (\bibinfo {year}
  {1968})}\BibitemShut {NoStop}%
\bibitem [{\citenamefont {Tyson}\ and\ \citenamefont
  {Othmer}(1978)}]{tyson1978dynamics}%
  \BibitemOpen
  \bibfield  {author} {\bibinfo {author} {\bibfnamefont {J.~J.}\ \bibnamefont
  {Tyson}}\ and\ \bibinfo {author} {\bibfnamefont {H.~G.}\ \bibnamefont
  {Othmer}},\ }\bibfield  {title} {\bibinfo {title} {The dynamics of feedback
  control circuits in biochemical pathways},\ }\href@noop {} {\bibfield
  {journal} {\bibinfo  {journal} {Progress in theoretical biology}\ }\textbf
  {\bibinfo {volume} {5}},\ \bibinfo {pages} {1} (\bibinfo {year}
  {1978})}\BibitemShut {NoStop}%
\bibitem [{\citenamefont {Fonkeu}\ \emph {et~al.}(2019)\citenamefont {Fonkeu},
  \citenamefont {Kraynyukova}, \citenamefont {Hafner}, \citenamefont {Kochen},
  \citenamefont {Sartori}, \citenamefont {Schuman},\ and\ \citenamefont
  {Tchumatchenko}}]{fonkeu2019mrna}%
  \BibitemOpen
  \bibfield  {author} {\bibinfo {author} {\bibfnamefont {Y.}~\bibnamefont
  {Fonkeu}}, \bibinfo {author} {\bibfnamefont {N.}~\bibnamefont {Kraynyukova}},
  \bibinfo {author} {\bibfnamefont {A.-S.}\ \bibnamefont {Hafner}}, \bibinfo
  {author} {\bibfnamefont {L.}~\bibnamefont {Kochen}}, \bibinfo {author}
  {\bibfnamefont {F.}~\bibnamefont {Sartori}}, \bibinfo {author} {\bibfnamefont
  {E.~M.}\ \bibnamefont {Schuman}},\ and\ \bibinfo {author} {\bibfnamefont
  {T.}~\bibnamefont {Tchumatchenko}},\ }\bibfield  {title} {\bibinfo {title}
  {How {mRNA} localization and protein synthesis sites influence dendritic
  protein distribution and dynamics},\ }\href@noop {} {\bibfield  {journal}
  {\bibinfo  {journal} {Neuron}\ }\textbf {\bibinfo {volume} {103}},\ \bibinfo
  {pages} {1109} (\bibinfo {year} {2019})}\BibitemShut {NoStop}%
\bibitem [{\citenamefont {Sturrock}\ \emph {et~al.}(2011)\citenamefont
  {Sturrock}, \citenamefont {Terry}, \citenamefont {Xirodimas}, \citenamefont
  {Thompson},\ and\ \citenamefont {Chaplain}}]{sturrock2011spatio}%
  \BibitemOpen
  \bibfield  {author} {\bibinfo {author} {\bibfnamefont {M.}~\bibnamefont
  {Sturrock}}, \bibinfo {author} {\bibfnamefont {A.~J.}\ \bibnamefont {Terry}},
  \bibinfo {author} {\bibfnamefont {D.~P.}\ \bibnamefont {Xirodimas}}, \bibinfo
  {author} {\bibfnamefont {A.~M.}\ \bibnamefont {Thompson}},\ and\ \bibinfo
  {author} {\bibfnamefont {M.~A.}\ \bibnamefont {Chaplain}},\ }\bibfield
  {title} {\bibinfo {title} {Spatio-temporal modelling of the hes1 and p53-mdm2
  intracellular signalling pathways},\ }\href@noop {} {\bibfield  {journal}
  {\bibinfo  {journal} {Journal of Theoretical Biology}\ }\textbf {\bibinfo
  {volume} {273}},\ \bibinfo {pages} {15} (\bibinfo {year} {2011})}\BibitemShut
  {NoStop}%
\bibitem [{\citenamefont {Gonze}\ and\ \citenamefont
  {Abou-Jaoud{\'e}}(2013)}]{gonze2013goodwin}%
  \BibitemOpen
  \bibfield  {author} {\bibinfo {author} {\bibfnamefont {D.}~\bibnamefont
  {Gonze}}\ and\ \bibinfo {author} {\bibfnamefont {W.}~\bibnamefont
  {Abou-Jaoud{\'e}}},\ }\bibfield  {title} {\bibinfo {title} {The {G}oodwin
  model: behind the {H}ill function},\ }\href@noop {} {\bibfield  {journal}
  {\bibinfo  {journal} {PloS one}\ }\textbf {\bibinfo {volume} {8}},\ \bibinfo
  {pages} {e69573} (\bibinfo {year} {2013})}\BibitemShut {NoStop}%
\bibitem [{\citenamefont {Erban}\ and\ \citenamefont
  {Chapman}(2020)}]{erban2020}%
  \BibitemOpen
  \bibfield  {author} {\bibinfo {author} {\bibfnamefont {R.}~\bibnamefont
  {Erban}}\ and\ \bibinfo {author} {\bibfnamefont {S.~J.}\ \bibnamefont
  {Chapman}},\ }\href@noop {} {\emph {\bibinfo {title} {Stochastic modelling of
  reaction--diffusion processes}}}\ (\bibinfo  {publisher} {Cambridge
  University Press},\ \bibinfo {year} {2020})\BibitemShut {NoStop}%
\end{thebibliography}%

\end{document}


\title{Supplemental Material:\\ Biological rhythms generated by a single activator-repressor loop with heterogeneity and diffusion.}

\author{Pablo Rojas}
\affiliation{ Theoretical Physics and Center for Interdisciplinary Nanostructure Science and Technology (CINSaT), University of Kassel, Kassel, Germany}
\author{Oreste Piro}%
\affiliation{Department of Ecology and Marine Resources, Institut Mediterrani d’Estudis Avançats, IMEDEA (CSIC–UIB), Balearic Islands, Spain}
\affiliation{Departament de F{\'i}sica, Universitat de les Illes Balears, Ctra. de Valldemossa, km 7.5, Palma de Mallorca E-07122, Spain}
\author{Martin E. Garcia}
\affiliation{ Theoretical Physics and Center for Interdisciplinary Nanostructure Science and Technology (CINSaT), University of Kassel, Kassel, Germany}%

\maketitle

\tableofcontents

\section{\label{sec:goodwin} Oscillations in Goodwin model }

The Goodwin model consists of a set of ordinary differential equations (ODEs), that represents a negative feedback loop in a chemical reaction network \cite{goodwin65,griffith68}. In this network, the protein at the end of the loop represses its own transcription. For each substance in the loop, the temporal evolution of its amount is described by production and degradation terms. In its simplest version, the only nonlinearity appears in the repression as a Hill function. The model assumes a well-mixed or compartmental condition. Eq. 1 and Eq. 2 from the main text are special cases of the more general $n$ reaction steps version, that reads as in Eq.\ref{eq:goodwin_n}. 

\begin{equation}
\label{eq:goodwin_n}
\begin{aligned}
	\frac{dx_1}{dt} & = \omega_{1} \frac{1}{1 + x_n^h} - \gamma_{1}x_1 \\
	\frac{dx_2}{dt} & = \omega_{2}x_1 - \gamma_{2}x_2 \\
	: &   \\
	\frac{dx_n}{dt} & = \omega_{n}x_{n-1} - \gamma_{n}x_n \\
\end{aligned}
\end{equation}

The secant condition, derived in \cite{tyson1978dynamics}, is the necessary condition that the exponent $h$ of the Hill function must fulfill to enable oscillatory solutions (Eq.\ref{eq:secant_condition}, Fig.1 in main text).
\begin{equation}
\label{eq:secant_condition}
h \geq \sec^n \left(\frac{\pi}{n} \right)  
\end{equation}

\section{\label{sec:deterministic} Deterministic model}

The modeled system is an extension of the Goodwin model to account for spatial heterogeneity, which transforms the set of ODEs into a set of partial differential equations (PDEs). Cell is represented by a symmetric spatial domain. For the sake of simplicity, no distinction between nucleus and cytoplasm transport properties is considered, i.e. homogenous diffusion coefficient. The membrane of the cell at the borders is impermeable.

The model first considers 2 species, mRNA and protein, whose spatial concentration are represented by  $m(\vec{r},t)$ and $p(\vec{r},t)$, respectively. Transcription of mRNA molecules and translation of proteins similarly to Goodwin model, but spatially restricted with functions $f_{GEN}(\vec{r})$ and $f_{RIB}(\vec{r})$, that represent the gene and ribosome locations, respectively. Linear degradation is considered to occur across the whole domain for both species, therefore it is equivalent to Goodwin model. We represented the molecular transport of mRNA and protein with Fickian diffusion. A parameter $p_{thresh}$ is used in the repression to control the scale of concentration of protein in the Hill function. The equations for the system are:
\begin{eqnarray}
\label{eq:deterministic2elem}
\dot{m}(\vec{r},t)   & = & \frac{\omega_m}{1+ \left(\dfrac{p(\vec{r},t) }{p_{thresh}} \right)^h} \, f_{GEN}(\vec{r}) -\gamma_m \, m(\vec{r},t)   + D_m \, \nabla^2 m(\vec{r},t)    \\
\dot{p}(\vec{r},t) & =& \omega_p \; f_{RIB}(\vec{r}) \, m(\vec{r},t)- \gamma_p \, p(\vec{r},t)   + D_p \nabla^2 \, p(\vec{r},t)    \nonumber 
\end{eqnarray}

This model represents our problem in the three dimensional space. To simplify our analysis we present our result with the reduction to one dimensional domain. Therefore, the equations read as:

\begin{eqnarray}
\label{eq:deterministic2elem1D}
\dot{m}(x,t)   & = & \frac{\omega_m}{1+    \left(\dfrac{p(x,t)  }{p_{thresh}} \right)^h} \, f_{GEN}(x) -\gamma_m \, m(x,t)  + D_m \, \nabla^2 m(x,t)   \nonumber \\
\dot{p}(x,t)   & = & \omega_p \; f_{RIB}(x) \, m(x,t) - \gamma_p \, p(x,t)  + D_p \nabla^2 \, p(x,t)      
\end{eqnarray}

The localization of gene and ribosomes is set to boxcar functions  $f_{GEN}(x)$ and $f_{RIB}(x)$ centered at $x_m$ and $x_p$, respectively, with radii $R_{m}$ and $R_{p}$.

\begin{equation}
\label{eq:fgen}
f_{GEN}(x) = 
    \begin{cases}       
        1 & \text{if } \; x \in [x_m - R_{m},x_m + R_{m} ]    \\
        0 & \text{else }
    \end{cases}
\end{equation}

\begin{equation}
\label{eq:frib}
f_{RIB}(x) = 
    \begin{cases}       
        1 & \text{if } \; x \in [x_p - R_{p},x_p + R_{p} ]    \\
        1 & \text{if } \; x \in [-x_p - R_{p},-x_p + R_{p} ]    \\
        0 & \text{else }
    \end{cases}
\end{equation}

The model as in Eq. \ref{eq:deterministic2elem1D} applies repression to the transcription of mRNA by comparing the values of $p(x,t)$ at each point $x$ with the respective parameter $p_{thresh}$. Due to the nonlinear nature of the Hill function, this differs from the well-mixed formulation in which the total amount is first computed and then compared to the respective parameter. We found these differences to lead to slight quantitative differences, only affecting the exact values for the onset of oscillations but preserving the structure of the solutions. Therefore, unless otherwise stated, the study shows results using Eq. \ref{eq:deterministic2elem1D}. However, for comparison with existing models we resorted to a formulation that compares the value of the repressing species, computed as the amount that is located within the gene location. We called this \emph{volumetric repression} (Eq. \ref{eq:deterministic2elem1Dvolum}).

\begin{eqnarray}
\label{eq:deterministic2elem1Dvolum}
\dot{m}(x,t)   & = & \frac{\omega_m}{1+    \left(\dfrac{P_{GEN}(t)  }{P_{thresh}} \right)^h} \, f_{GEN}(x) -\gamma_m \, m(x,t)  + D_m \, \nabla^2 m(x,t)   \nonumber \\
\dot{p}(x,t)   & = & \omega_p \; f_{RIB}(x) \, m(x,t) - \gamma_p \, p(x,t)  + D_p \nabla^2 \, p(x,t)      ,
\end{eqnarray}

where $ P_{GEN}(t) $ is the amount of $ p(x,t) $ integrated in the gene location (Eq. \ref{eq:Pgen_integral_value}). 

\begin{equation}
\label{eq:Pgen_integral_value}
	P_{GEN}(t)  =  \int_\Omega p(x,t) f_{GEN}(x) dx, 
\end{equation}
where $\Omega$ is the cell volume. 

The solution of the PDEs were obtained numerically using second order finite differences in the spatial domain, and Runge-Kutta-4 in the time domain.

\section{\label{sec:three_molecules} Feedback loop with three species }

The extension of our model to 3 participating species is done similarly to 2 species. The equations for 3 participating species that is the corresponding to Eq. \ref{eq:deterministic2elem1D} for 2 species, are shown in Eq. \ref{eq:3elem}.

\begin{eqnarray}
\label{eq:3elem}
\dot{m}(x,t)   & = & \frac{\omega_m}{1+   \left(\dfrac{r(x,t)  }{r_{thresh}} \right)^h} \, f_{GEN}(x) -\gamma_m \, m(x,t)  + D_m \, \nabla^2 m(x,t)   \nonumber \\
\dot{p}(x,t)   & = & \omega_p \; f_{RIB}(x) \, m(x,t)- \gamma_p \, p(x,t)  + D_p \nabla^2 \, p(x,t)        \\
\dot{r}(x,t)   & = & \omega_r \; f_{REP}(x) \, p(x,t)- \gamma_p \, r(x,t)   + D_r \nabla^2 \, r(x,t)    \nonumber 
\end{eqnarray}

The function $f_{REP}(x)$ represents the site where the production of the repressor $r$ takes place. To simplify the analysis, $f_{REP}(x)$ is chosen to be identical to $f_{RIB}(x)$ (Eq.\ref{eq:frep}), so that only one distance remains as a parameter of the problem.
 
\begin{equation}
\label{eq:frep}
f_{REP}(x) = f_{RIB}(x) = 
    \begin{cases}       
        1 & \text{if } \; x \in [x_p - R_{p},x_p + R_{p} ]    \\
        1 & \text{if } \; x \in [-x_p - R_{p},-x_p + R_{p} ]    \\        
        0 & \text{else }
    \end{cases}
\end{equation}

The model with 3 species have a better comparison with ODE models when the formulation with volumetric repression is used. Under this framework, Eq. \ref{eq:3elem} is modified as:

\begin{eqnarray}
\label{eq:3elem_volum}
\dot{m}(x,t)   & = & \frac{\omega_m}{1+   \left(\dfrac{R_{GEN}(t)  }{R_{thresh}} \right)^h} \, f_{GEN}(x) -\gamma_m \, m(x,t)  + D_m \, \nabla^2 m(x,t)   \nonumber \\
\dot{p}(x,t)   & = & \omega_p \; f_{RIB}(x) \, m(x,t) - \gamma_p \, p(x,t)  + D_p \nabla^2 \, p(x,t)        \\
\dot{r}(x,t)   & = & \omega_r \; f_{REP}(x) \, p(x,t) - \gamma_p \, r(x,t)   + D_r \nabla^2 \, r(x,t)    \nonumber 
\end{eqnarray}

where 

\begin{equation}
\label{eq:Rgen_integral_value}
	R_{GEN}(t)  =  \int_\Omega r(x,t) f_{GEN}(x) dx, 
\end{equation}

In the examples shown for 3 species, we used volumetric repression.

\begin{figure}[h]
    \includegraphics[width=6.6in]{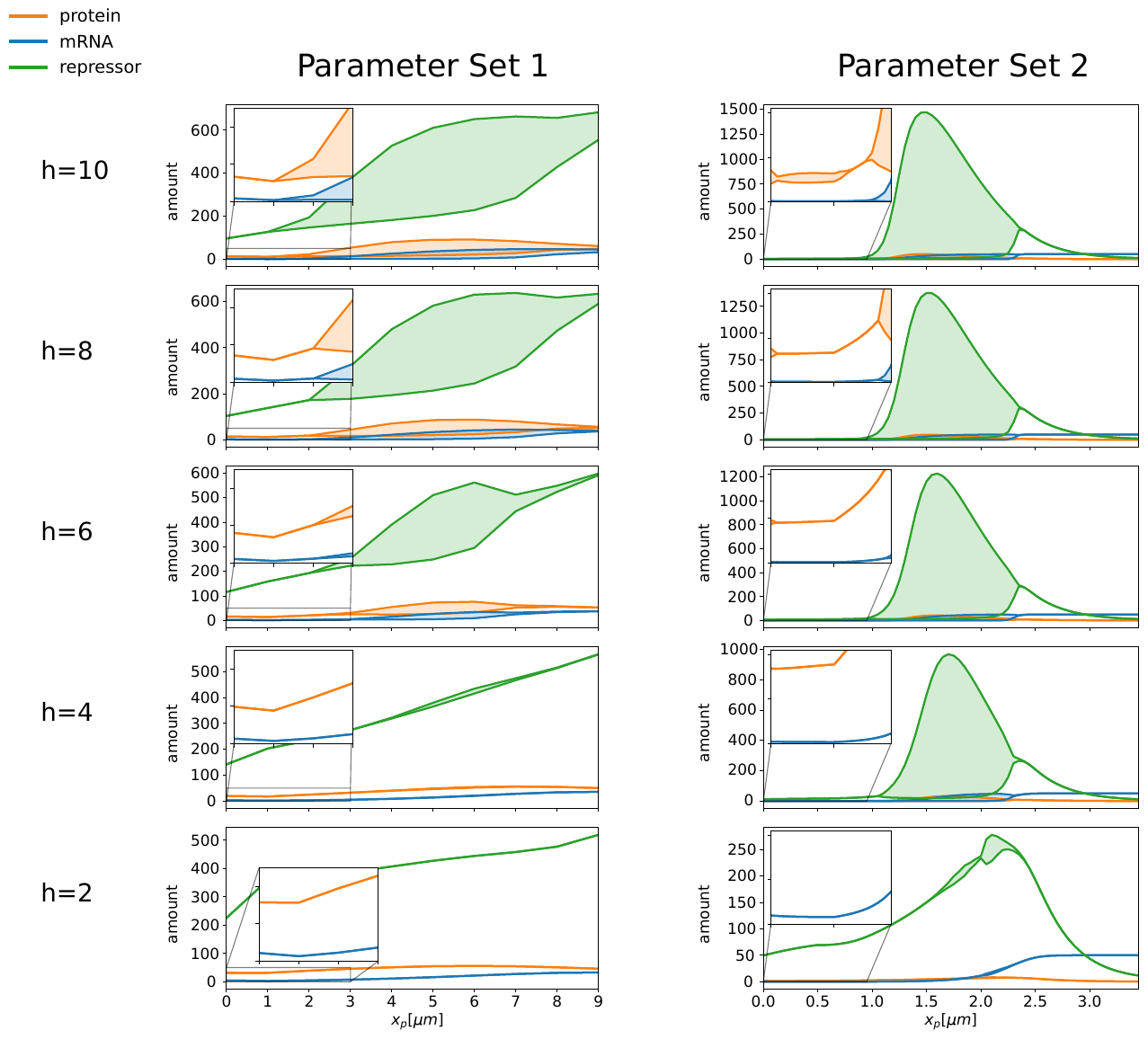}
    \caption{ Amplitude of oscillations in an extension of the model for 3 species. Left panel: Parameter Set 1 described in the main text. Right panel: Parameter Set 2. Lines indicate minimum and maximum of the trajectories in the long term approximation. Shaded areas between minimum and maximum indicate amplitude of oscillations. Parameter Set 1 only displays oscillations when there is a separation between the sources. Parameter Set 2 shows oscillations when sources are co-localized (as predicted by the well-mixed model) and when sources are separated. Notice the intermediate region when no oscillation is present. Oscillations in the separation case present higher amplitude.  }
    \label{fig:3elem_loop}
\end{figure}

The problem is solved for 2 sets of parameters. In the Parameter Set 1, parameters for the transport of substances are selected according to the ones reviewed and used in \citep{fonkeu2019mrna,sturrock2011spatio}. These parameters represent experimental measurement of diffusion coefficients and kinetic rates for mRNA and protein. Since in this work reviews other sources of experimental data and reports a wide variability on the measurements, we simplified the choice of parameters by using the corresponding order of magnitude. In the Parameter Set 2, we selected kinetic parameters that would generate oscillations in the well-mixed Goodwin Model, inspired in Ref. \citep{gonze2013goodwin}, in order to induce an equivalence to the corresponding well-mixed version under low values of $x_p$. Full set of parameters in Table \ref{tab:param3elem}.

\section{\label{sec:stochastic} Stochastic model }

We constructed a stochastic model of the coupled reaction and diffusion of the systems, by treating each of the molecules as an individual agent, as shown in Fig. \ref{fig:stochastic_scheme}. The Eq. \ref{eq:SSA} represents the sets of equations that represent the creation, degradation and displacements of the molecules. In these equations, $\phi$ denotes chemical species that are of no interest, and is used in order to represent degradation and creation of molecules. To solve this system, we used the stochastic simulation algorithm (SSA) described in Ref. \cite{erban2020}. 
\begin{eqnarray}
\label{eq:SSA}
mRNA     & \rightarrow   & \phi        \nonumber  \\
protein  & \rightarrow   & \phi         \nonumber \\
\phi     & \rightarrow  & mRNA           \\
\phi     & \rightarrow    & protein     \nonumber \\
x_i(t+\Delta t) & =&  x_i(t ) + \sqrt{2D \Delta t} \xi , \nonumber 
\end{eqnarray}
In our SSA, we used a time step $\Delta t$, small enough so that the probability of a certain reaction ocurring during this interval of time is $k \Delta t$, $k$ being its reaction rate. We therefore generate random numbers at each time step to determine if an individual molecule is degraded. We also determine whether a new molecule of a certain species must be created, in which case it will be positioned at a random location within its own creation zone. During the same interval, an existing molecule can update its position by determining its stochastic displacement $\sqrt{2D \Delta t} \xi$, where $ \xi $ is a number drawn from a Gaussian distribution. We therefore determine, at each time step, whether an individual molecule is degraded, created and where its new position is, if corresponding (Alg. \ref{alg:SSA_steps}). The reaction rates are determined using the equivalent expresions from the deterministic model (Eq. \ref{eq:reaction_rates_SSA}).
\begin{eqnarray}
\label{eq:reaction_rates_SSA}
k_{degradation \, mRNA}     & =   &  \gamma_m        \nonumber  \\
k_{degradation \, protein}  & =   &  \gamma_p         \nonumber \\
k_{creation \, mRNA}     & =   & \frac{\omega_m}{1+    \left(\frac{protein_{transcription \; zone}  }{protein_{thresh}} \right)^h}       \\
k_{creation \, protein}  & =   &  \omega_p \quad {mRNA}_{translation \, zone} \,        \nonumber 
\end{eqnarray}

\begin{figure}
\begin{algorithm}[H]
\caption{Stochastic model steps. }\label{alg:SSA_steps}
\begin{algorithmic}
\State set initial condition and parameters
\While{$t \leq t_{max}$} 
	\State determine mRNA (protein) molecules in translation (transcription) zone
	\State compute reaction rates and probabilities
	\For{each molecule}
		\State generate random numbers	
		\State compute displacements
		\State update positions
	\EndFor
	\For{each molecule}
		\State generate random number
		\If{$random \quad number \leq probability \quad degradation$}
		    \State eliminate molecule
    		\EndIf
	\EndFor
	\For{each species (mRNA and Protein)}
		\State generate random number
		\If{$random \quad number \leq probability \quad creation$}
		    \State insert new molecule in random position within production zone
    		\EndIf
	\EndFor
\EndWhile
\end{algorithmic}
\end{algorithm}
\end{figure}

\begin{figure}[h!]
    \includegraphics[width=6.0in]{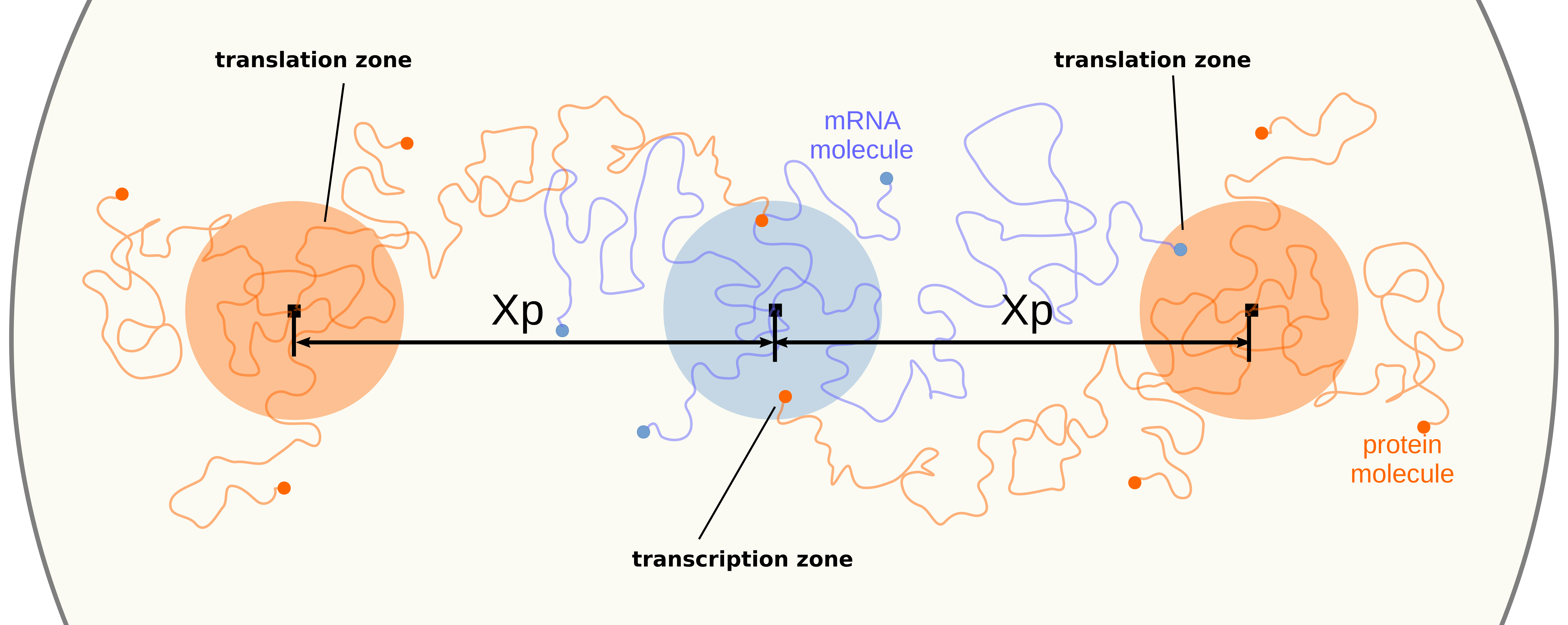}
    \caption{ Scheme of the stochastic model. Molecules are modeled as individual particles. They are created either at gene (transcription zone) or ribosome (translation zone), similarly to deterministic model. To create molecules, the stochastic simulation algorithm (SSA) is implemented with a propensity function based on the number of molecules of the other species in their own production zone. Molecules travel following the Langevin equation, and are reflected at the borders of the cell. The results shown in Fig. 4 in the main text where obtained with a 1D version of this model. }
    \label{fig:stochastic_scheme}
\end{figure}

\section{\label{sec:relaxation} Derivation of relaxation times }

\subsection*{Reduced problem}
We first consider a symmetric one dimensional domain, in which for $t\geq0$ the production of a substance whose concentration is $c(x,t)$ is done at a constant rate $\omega$ in a punctual source centered in the interval $[-L,L]$, represented by $\delta(x)$. At $t=0$, $c(x,t=0)=0$ in the whole spatial domain. This substance is subjected to Fickian diffusion (with a constant diffusion constant $D$), and degradation (with $\gamma$ as degradation constant). The walls of the domain $[-L,L]$ are impermeable. For $t<0$ the system is at its equilibrium state $c(x)=0$ since no source is available, and this equilibrium is perturbed by the sudden appearance of the punctual source. This scenario is an idealized cartoon of the instant in which hypothetical mRNA (protein) initiates transcription (translation) inside a cell, without prior concentration. The new concentration as approaches equilibrium at long times. We will derived a characteristic time that describes the relaxation of the concentration at a certain spatial point, from its initial condition $c=0$ to the new established profile $c=c_{t \rightarrow \infty}$. The equation for the reduced problem reads:

\begin{equation}
\label{eq:finite_problem}
\left\{
\begin{aligned}
	\frac{\partial c}{\partial t} 			&= D  \nabla^2 c - \gamma c + \omega  \delta (x) 
													& \quad \forall x \mbox{ in } [-L,L]  \\
	\left.\frac{\partial c}{\partial x} \right|_L 	&= \left.\frac{\partial c}{\partial x} \right|_{-L} = 0  \\
	c(x,0) &= 0 & \forall x \mbox{ in } [-L,L]\\
\end{aligned} 
\right.
\end{equation}

Since the problem is symmetric, we can rewrite the system for the positive sub-interval, by transforming the production term into a boundary condition $\left( \left.\frac{\partial c}{\partial x} \right|_0 = -\frac{\omega}{2D}\right)$.

\begin{equation}
\left\{
\begin{aligned}
	\frac{\partial c}{\partial t} 			&= D  \nabla^2 c - \gamma c  
													& \quad \forall x \mbox{ in }[0,L]  \\
	\left.\frac{\partial c}{\partial x} \right|_0 	&= -\frac{\omega}{2D}  \\
	\left.\frac{\partial c}{\partial x} \right|_L 	&= 0  \\
	c(x,0) &= 0 & \forall x \mbox{ in }[0,L]\\
\end{aligned} 
\right.
\end{equation}

The solution of the problem is simpler if we assume the limit $L \rightarrow \infty$. Since our purpose is to derive estimates analytically, this assumption proves to be practical and does not affect the conclusions. For $L \rightarrow \infty$ the problem reads:

\begin{equation}
\label{eq:infinite_symm_problem}
\left\{
\begin{aligned}
	\frac{\partial c}{\partial t} 			&= D  \nabla^2 c - \gamma c  
													& \quad \forall x \mbox{ in }[0,\infty)  \\
	\left.\frac{\partial c}{\partial x} \right|_0 	&= -\frac{\omega}{2D}  \\
	\left.\frac{\partial c}{\partial x} \right|_{x \rightarrow \infty} 	&= 0  \\
	c(x,0) &= 0 & \forall x \mbox{ in }[0,\infty)\\
\end{aligned} 
\right.
\end{equation}

The solution to the steady state problem, i.e. $ \frac{\partial c}{\partial t} = 0$ is asymptotically approached by the transient solution at long times $t \rightarrow \infty$. The steady state solution becomes:
\begin{equation}
\label{eq:steady_state_solution}
c_{t \rightarrow \infty}= \frac{\omega}{2 \sqrt{D \gamma}} e^{-\sqrt{\frac{\gamma}{D}} x} 
\end{equation}

To obtain the transient solution, we will work with the Laplace transform $u(x,s) = \mathcal{L}[c(x,t)]$. The resulting differential equation is:

\begin{equation}
\label{eq:laplace_ODE}
	\frac{\partial^2 u}{\partial x^2} - \left( \frac{s+\gamma}{D} \right) u = 0 
\end{equation}
and we can propose a solution of the form:
\begin{equation*}
	u(x,s)= C_1 e^{-\sqrt{\frac{s +\gamma}{D}} x} 
\end{equation*}
 
By applying the boundary condition at $x=0$:

\begin{equation*}
	\frac{\partial u}{\partial x}=   \mathcal{L} \left[ \frac{\partial c}{\partial x}\right]  
\end{equation*}
\begin{equation*}
	C_1 \left( -\sqrt{\frac{s +\gamma}{D}} \right) = -\frac{\omega}{2D \, s}
\end{equation*}
\begin{equation*}
	\Rightarrow C_1  =  \frac{\omega}{2 \sqrt{D} \, s \, \sqrt{s +\gamma}} 
\end{equation*}

Then, the solution for $u$ is: 
\begin{equation}
\label{eq:solution_u}
	u(x,s)= \frac{\omega}{2 \sqrt{D} \, s \, \sqrt{s +\gamma}} e^{-\sqrt{\frac{s +\gamma}{D}} x} 
\end{equation}

To obtain $c(x,t)$, it is necessary to anti-transform the solution by $c(x,t) = \mathcal{L}^{-1}[ u(x,s)]$. 

\begin{equation}
\label{eq:antitransform_c}
	c(x,t)= \mathcal{L}^{-1} \left[ \frac{\omega}{2 \sqrt{D} \, s \, \sqrt{s +\gamma}} e^{-\sqrt{\frac{s +\gamma}{D}} x} \right]
\end{equation}

It is possible to antitransform first the factors $\mathcal{L}^{-1} \left[ \frac{1}{s} \right] $ and $\mathcal{L}^{-1} \left[ \frac{e^{-\sqrt{\frac{s +\gamma}{D}} x}}{\sqrt{s +\gamma}} \right] $, and use these results to arrive to the solution of $c(x,t)$.

\begin{equation}
\label{eq:solution_analytic_with_integral}
	c(x,t)=  \frac{\omega}{2 \sqrt{D \pi } } \int_{0}^{t} \frac{e^{- \left( \gamma y + \frac{x^2}{4Dy} \right)}}{\sqrt{y}} dy
\end{equation}

Solving the integral:

\begin{equation}
c(x,t) = \frac{\omega}{4\sqrt{D \gamma}} \left[ e^{-\sqrt{\frac{\gamma}{D}} x}  \erfc\left( \frac{x}{2\sqrt{D t}} - \sqrt{\gamma t} \right)   - e^{\sqrt{\frac{\gamma}{D}} x}  \erfc\left( \frac{x}{2\sqrt{D t}} + \sqrt{\gamma t} \right) \right] 
\label{solution_analytic}
\end{equation}

This solution is displayed in Fig.\ref{fig:delays_analytic}a for different times. 

\subsection*{Relaxation time}
The time at which $\frac{\partial c}{\partial t}$ maximizes can be obtained by the solution of:

\begin{equation}
\label{eq:maximal_c_rate}
	\frac{\partial }{\partial t} \left( \frac{\omega}{2 \sqrt{D \pi } }  \frac{e^{- \left( \gamma t + \frac{x^2}{4Dt} \right)}}{\sqrt{t}} \right) = 0 
\end{equation}

From which we can obtain:

\begin{equation*}
	\frac{\partial }{\partial t} \left(  \frac{e^{- \left( \gamma t + \frac{x^2}{4Dt} \right)}}{\sqrt{t}} \right) = \left( - \frac{e^{- \left( \gamma t + \frac{x^2}{4Dt} \right)}}{\sqrt{t}} \right) \frac{1 }{t^{3/2} } \left(      \gamma t - \frac{x^2}{4D t} + \frac{ 1 }{2} \right) 	 = 0
\end{equation*}

Therefore, the \emph{relaxation time} $t_{R}$ results as a solution of:
\begin{equation}
\label{eq:equation_relax_time}
t_{R}^2 +\frac{t_{R}}{2\gamma} - \frac{x^2}{4D\gamma} = 0
\end{equation}

\begin{figure}[h]
    \includegraphics[width=6.6in]{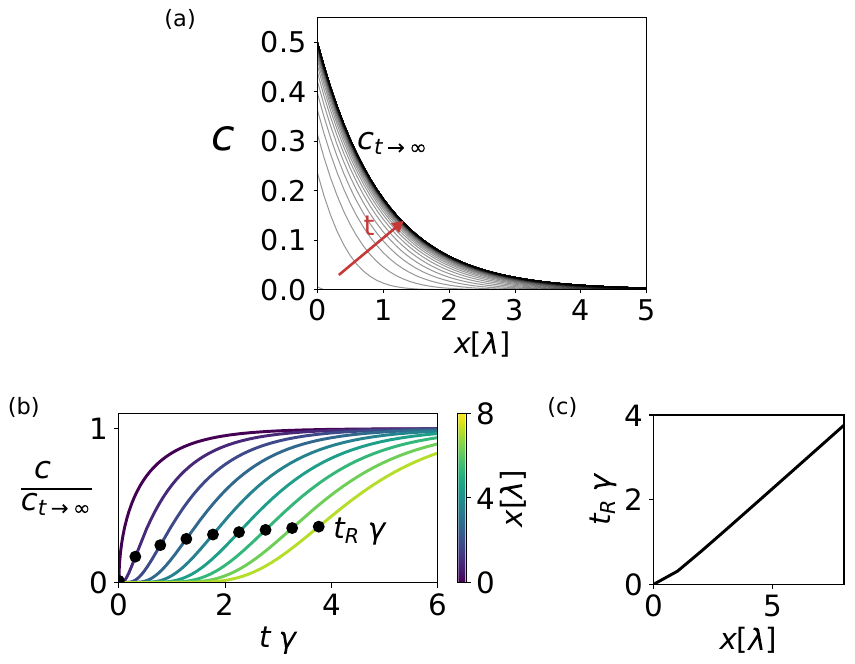}
    \caption{ Relaxation times in the reduced problem.  (a) Concentration of the substance $c$ as a function of $x$ at different times. The spatial profile converges asymptotically to the steady state. (b) Normalized concentration as a function of re-scaled time. Dots represent relaxation times, defined as the times in which the rate of increase of concentration is maximal. (c) Analytical solution for relaxation times show asymptotically linear dependence with distance to the source. }
    \label{fig:delays_analytic}
\end{figure}

Since we are interested in the interval $t\geq 0$, one of the solutions is ruled out and we obtain:
\begin{equation}
\label{eq:solution_relax_time}
t_{R} = \frac{1}{4\gamma}   \left(  \sqrt{1 + 4  \left(\frac{x}{\lambda}\right)^2 } -1  \right)
\end{equation}

where $ \lambda  = D/\gamma$ is the characteristic length. For values of $x/\gamma$ large enough, $t_{R}$ approaches a linear dependence on $x$ (Fig.\ref{fig:delays_analytic}b-c) (notice the difference with the $x^2/D$ dependence of time-scales in purely diffusive systems).

\subsection*{Dimensionless version of the reduced problem}

A change of variables that render the reduced problem as dimensionless can help the relevant parameters. The change of variables:

\begin{center}
$
\begin{array}{rllcrl}
	x^\ast & = \frac{x}{\lambda} & = \frac{x}{\sqrt{\frac{D}{\gamma}}} & \Rightarrow & \; x & =  x^* \sqrt{\frac{D}{\gamma}} \\
	t^* & = \frac{2 t \sqrt{D \gamma}}{x}  & = \frac{2 t }{ \frac{x}{\lambda \gamma  } } & \Rightarrow & \; t & =  \frac{t^* x^*}{2 \gamma}
\end{array}
$
\end{center}

yields the steady state solution ($t\rightarrow \infty$)  re-written as:

\begin{center}
$
\begin{array}{rlcrl}
	c^* & = \frac{c}{\frac{\omega}{2\sqrt{D \gamma}}} &  &  &  \\
	c_{t\rightarrow \infty}(x) & =  \frac{\omega}{2\sqrt{D \gamma}} e^{-\sqrt{\frac{\gamma}{D}} x} & \Rightarrow & \; c_{t\rightarrow \infty}^*(x^*) & =  e^{-x^*}
\end{array}
$
\end{center}

For the time-varying solution the dimensionless version yields:
\begin{equation}
\label{eq:dimensionless_solution}
c^*(x^*,t^*) = \frac{1}{2} \left[ e^{-x^*}  \erfc\left( \frac{1}{\sqrt{2}} \left[ \sqrt{\frac{x^*}{t^*}} - \sqrt{x^* t^* } \right] \right)   - e^{x^*}  \erfc\left( \frac{1}{\sqrt{2}} \left[ \sqrt{\frac{x^*}{t^*}} + \sqrt{x^* t^* } \right] \right) \right] 
\end{equation}

therefore, the problem scales with $x$ proportional to lambda $\lambda$, but in time $t$ it scales with $x/\lambda$ and inversely to $\gamma$.

\subsection*{Linear dependence of delay on distance in  oscillatory solutions}

\begin{figure}[h]
    \includegraphics[width=6.6in]{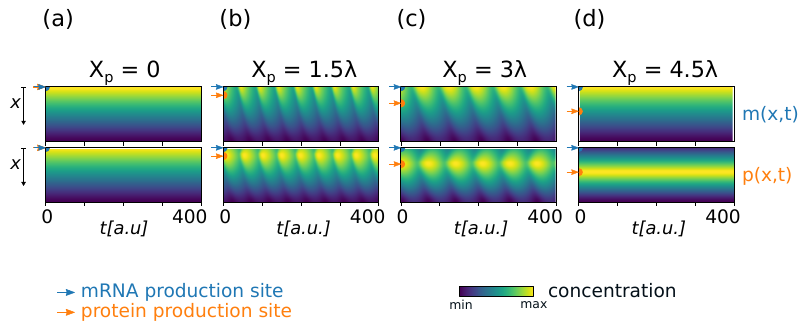}
    \caption{ Propagation of concentration gradients arising from spatial separation of the sources. (a) Distribution of the species $m(x,t)$ and $p(x,t)$ in space and time, between center and borders of the cell, for different separation of sources. Concentrations are shown in normalized logarithmic scale. The two displayed oscillatory solutions ($x_p = 1.5 \lambda$ and  $x_p = 3 \lambda$) show different periods of oscillation, but the same dependence of diffusion-induced delays with distance (slope of the bright regions). The parameters for this figure correspond to the ones in Fig.4 in main text. $h= 10 $, $\lambda = \lambda_m = \lambda_p =1$, $ D_m = D_p = 0.1$ , $ \gamma_m = \gamma_p = 0.1 $, $ R_{m} = R_{p} = 0.5$, $ R_{cell} = 10$ , $\omega_m = \omega_p = 10 $  } 
    \label{fig:heatmap_space_time}
\end{figure}

The spatio-temporal distribution of $m(x,t)$ and $p(x,t)$ in oscillatory solutions (Fig.\ref{fig:heatmap_space_time}) described in the main text, shows an almost constant speed of propagation , in agreement with the derived dependence on $x$. 
For sufficiently small or sufficiently large separation of the sources, the spatial distribution of $m(x,t)$ and $p(x,t)$ remain constant over time (Fig.\ref{fig:heatmap_space_time}a,d).
However, for intermediate separations, these distributions display oscillatory behavior in time (Fig. \ref{fig:heatmap_space_time}b,d).
The period of oscillations increases as the distance between sources increases. Propagation in the spatial distribution of species are recognizable as the slanted brighter regions. The slope of the regions denotes the constant propagation speed.

\section{\label{sec:params} Parameters }

Parameters used in the models with two species are given in arbitrary units (Table \ref{tab:param2elem}), with the exception that the amounts can be interpreted in molecules. Parameters in models for three species are selected based on previous work with their corresponding units (Table \ref{tab:param3elem}). Nevertheless, parameters in models with two species could be interpreted as having the same units as the models for three species, i.e. $\mu m$ for sizes and distances, $h$ for time and their corresponding derived units. 

\begin{table}[h!]
\caption{\label{tab:param2elem}
Parameters used in the models with 2 species. Values are given in arbitrary units. }
\begin{ruledtabular}
\begin{tabular}{lcccccccccc}
 &$ D_{m}$  & $D_{p}$ &  $ \gamma_{m}$  & $ \gamma_{p}$& $ \omega_{m}$  & $ \omega_{p}$&$R_{cell}$ &$ R_{m}$  & $R_{p}$& $p_{thresh}$ \\
\hline
Fig. 2-3& 0.1 & 0.1 & 0.1 & 0.1 & 10.0 & 20.0 & 10.0 & 0.5 & 0.5 & 0.1\\
Fig. 4  & 0.1 & 0.1 & 0.1 & 0.1 & 10.0 & 10.0 & 10.0 & 0.5 & 0.5 & 0.1\\
\end{tabular}
\end{ruledtabular}
\end{table}

\begin{table}[h!]
\caption{\label{tab:param3elem}
Parameters used in the models with 3 species. }
\begin{ruledtabular}
\begin{tabular}{lcccccccccccccc}
 &$ D_{m}$  & $D_{p}$ &$ D_{r}$ &  $ \gamma_{m}$  & $ \gamma_{p}$ &  $ \gamma_{r}$ & $ \omega_{m}$  & $ \omega_{p}$ & $ \omega_{r}$  &$R_{cell}$ &$ R_{m}$  & $R_{p}$  &$ R_{r}$ & $r_{thresh}$ \\
 &$  \left(\frac{\mu m ^2}{h}\right)$  & $  \left(\frac{\mu m ^2}{h}\right)$ &$  \left(\frac{\mu m ^2}{h}\right)$ &  $  \left(\frac{1}{h}\right)$   & $  \left(\frac{1}{h}\right)$  &  $  \left(\frac{1}{h}\right)$  & $  \left(\frac{mol.}{h\,\mu m}\right)$   & $  \left(\frac{mol.}{h\,\mu m}\right)$ & $  \left(\frac{mol.}{h\,\mu m}\right)$  &$ (\mu m) $ & $ (\mu m) $   & $ (\mu m) $   & $ (\mu m) $ & $ (mol.) $  \\
\hline
Param. Set 1 & 10.0 & 10.0 & 10.0 & 1.0 & 1.0 & 1.0 & 50.0 & 100.0 & 100.0 & 10.0 & 0.5 & 0.5 & 0.5 & 2.0\\
Param. Set 2 & 0.01 & 0.01 & 0.01 & 0.1 & 0.1 & 0.1 & 5.0  & 10.0  & 10.0  & 4.0  & 0.5 & 0.5 & 0.5 & 1.0\\
\end{tabular}
\end{ruledtabular}
\end{table}

\bibliography{supp_text}